\documentclass[aps,prb,twocolumn,notitlepage,footinbib]{revtex4-1}
\usepackage[a4paper,top=1.5cm,bottom=1.5cm,left=1.75cm,right=1.75cm]{geometry}
\usepackage{amsmath}
\usepackage{amssymb}
\usepackage{amsfonts}
\usepackage{hyperref}
\usepackage{mathptmx}	                     
\DeclareSymbolFont{newfont}{OML}{cmm}{m}{it} 
\DeclareMathSymbol{\eps}{3}{newfont}{15}     
\usepackage[utf8]{inputenc}
\usepackage[T1]{fontenc}
\usepackage{graphicx}
\DeclareGraphicsExtensions{.eps, .pdf, .jpg, .png}
\usepackage{enumerate}
\usepackage{bm}
\usepackage{bbold}
\usepackage{color}
\def\text#1{{\mbox{\tiny #1}}}
\def\point#1#2{{\sf #1}_{\mbox{\tiny #2}}}
\def\gauge{{\sf U(1)_{\text{N}}}}
\def\whG{\widehat{G}}
\def\whg{\widehat{g}}
\def\whT{\widehat{T}}
\def\whU{\widehat{U}}
\def\whSigma{\widehat{\Sigma}}
\def\whDelta{\widehat{\Delta}}
\def\dangle#1{\frac{d\Omega_{#1}}{4\pi}}
\newcommand{\sgn}{\mbox{sgn}}
\renewcommand{\Im}{\mbox{Im\,}}
\renewcommand{\Re}{\mbox{Re\,}}
\newcommand{\whmfG}{\widehat{\mathfrak{G}}}
\newcommand{\whmfGr}{\widehat{\mathfrak{G}}^{\text{R}}}

\def\vA{{\bf A}}
\def\vJ{{\bf J}}
\def\vE{{\bf E}}
\def\vp{{\bf p}}
\def\vv{{\bf v}}
\def\vq{{\bf q}}
\def\vr{{\bf r}}

\def\sml{{\bf\textsf{s}}}
\newcommand{\grad}{\pmb\nabla}
\newcommand{\be}{\begin{equation}}
\newcommand{\ee}{\end{equation}}
\newcommand{\ber}{\begin{eqnarray}}
\newcommand{\eer}{\end{eqnarray}}

\def\hone{\hat{1}}
\def\tone{\widehat{1}}
\def\tx{{\widehat{\tau}_1}}
\def\ty{{\widehat{\tau}_2}}
\def\tz{{\widehat{\tau}_3}}
\newcommand{\whtaux}{{\widehat{\tau}_1}}
\newcommand{\whtauy}{{\widehat{\tau}_2}}
\newcommand{\whtauz}{{\widehat{\tau}_3}}
\newcommand{\whone}{\widehat{1}}

\def\commutator#1#2{\mbox{$\displaystyle\,\left[\,#1\,,\,#2\,\right]$}}
\def\anticommutator#1#2{\mbox{$\displaystyle\left\{\,#1\,,\,#2\,\right\}$}}
\newcommand{\kb}{k_{\text{B}}}
\def\Tr#1{\mbox{Tr}\big\{#1\big\}}
\def\nicefrac#1#2{\genfrac{}{}{}{1}{#1}{#2}}
\def\ns{\negthickspace}

\def\SigN{\Sigma_{\text{N}}}
\def\SigS{\Sigma_{\text{S}}}
\def\DelN{\Delta_{\text{N}}}
\def\DelS{\Delta_{\text{S}}}
\def\whSigN{\whSigma_{\text{N}}}
\def\whSigS{\whSigma_{\text{S}}}
\def\nN{c_{\text{N}}}
\def\nS{c_{\text{S}}}
\def\uN{u_{\text{N}}}
\def\uS{u_{\text{S}}}
\def\vN{v_{\text{N}}}
\def\bvN{\bar{v}_{\text{N}}}
\def\bvNl{\bar{v}_{\text{N}l}}
\def\buN{\bar{u}_{\text{N}}}
\def\buS{\bar{u}_{\text{S}}}
\def\buSl{\bar{u}_{\text{S}l}}

\def\deltaN{\delta_{\text{N}}}
\def\deltaNl{\delta_{\text{N}_l}}

\def\deltaSl{\delta_{\text{S}_l}}
\def\GammaN{\Gamma_{\text{N}}}
\def\GammaS{\Gamma_{\text{S}}}
\def\alphaS{\alpha_{\text{S}}}
\def\alphaN{\alpha_{\text{N}}}

\def\TN{\whT_{\text{N}}}
\def\TS{\whT_{\text{S}}}
\def\tauN{\tau_{\text{N}}}
\def\tauS{\tau_{\text{S}}}
\def\tauSn{\bar{\tau}_{\text{S}_{\text{n}}}}
\def\omegaD{\omega_{\text{D}}}

\def\bepsR{\bar\varepsilon^{\text{R}}}
\def\bepsA{\bar\varepsilon^{\text{A}}}
\def\bepsRA{\bar\varepsilon^{\text{R,A}}}
\def\dpR{d_{+}^{\text{R}}}
\def\dpA{d_{+}^{\text{A}}}
\def\dpa{d_{+}^{\mbox{\footnotesize a}}}
\def\tauSR{\tilde\tau_{\text{S}}^{\text{R}}}
\def\tauSA{\tilde\tau_{\text{S}}^{\text{A}}}
\def\tauSRA{\tilde\tau_{\text{S}}^{\text{R,A}}}
\def\DeltaRA{\tilde\Delta^{\text{R,A}}}
\setcitestyle{numbers,square}
\begin{document}
\title{Microwave Response of Superconductors with Paramagnetic Impurities}
\author{Mehdi Zarea}
\email{zarea.mehdi@gmail.com}
\affiliation{Hearne Institute of Theoretical Physics, 
             Louisiana State University,
             Baton Rouge LA 70803 USA}
\author{J. A. Sauls}
\email{sauls@lsu.edu}
\affiliation{Hearne Institute of Theoretical Physics, 
             Louisiana State University,
             Baton Rouge LA 70803 USA}
\date{\today}

\begin{abstract}
We develop theoretical methods to predict the effects of paramagnetic impurities on the microwave response of conventional spin-singlet superconductors.
Our focus is on superconducting devices and resonators with \emph{low concentrations} of impurities and \emph{exchange interactions} with conduction electrons.
We connect the sub-gap quasiparticle spectrum generated by pair-breaking to the frequency and temperature dependence of the conductivity for superconductors operating at microwave frequencies.
We report theoretical results for superconducting device performance - dissipation, quality factor and frequency shift anomalies - based on self-consistent calculations of the current response and penetration of the electromagnetic field at the vacuum-superconducting interface. 
Key results include the prediction of a non-monotonic anomaly in the low-frequency superfluid fraction and penetration depth at very low temperatures related to the sub-gap quasiparticle spectrum.
Dissipation of microwave power is predicted from intra- and inter- impurity band transitions at GHz frequencies at low temperatures, including a physical mechanism responsible for residual resistance.
We predict anomalies in the resonant frequency, $f(T)$, and quality factor, $Q(T)$, of high-Q SRF cavities operating in the GHz range at low-temperatures that are sensitive to non-magnetic and paramagnetic impurity disorder.
\end{abstract}

\maketitle

\section{Introduction}\label{sec-Introduction}
\vspace*{-3mm}

A known source of decoherence for superconducting resonators and qubits is dissipation from quasiparticles under nonequilibrium conditions. Indeed quasiparticle dissipation is reported to be as important as all other loss channels in transmon qubits when coherence times exceed $200\,\mu\mbox{sec}$~\cite{ser18}.
Potential sources of a significant population of quasiparticles leading to decoherence in superconducting devices range from high current densities and multi-photon pair breaking~\cite{dev14,cho25}, energy deposition from background radiation and cosmic rays~\cite{car21,car23}, to sub-gap of Andreev bound states generated by surface defects~\cite{deg20} or pair-breaking by TLS impurities~\cite{he25}.

The effects of impurity disorder on superconductivity was studied extensively soon after the development of the microscopic theory of superconductivity~\cite{bar57} by many authors~\cite{and59,abr59b,abr61,mak63}.
For recent studies related to microwave resonators and SRF cavities see Refs.~\cite{fom11,kha12,sau22,uek25,zar23,uek24,leb25}.
In this report we focus on the role and impact of dilute concentrations of paramagnetic impurities on the microwave response of superconducting circuits and resonators.

\vspace*{-5mm}
\subsection{Background}
\vspace*{-3mm}

The exchange coupling of electrons to paramagnetic impurities is well known to break Cooper pairs, leading to un-paired quasiparticle states below the continuum threshold (``gap edge'') of quasiparticle excitations.
Abrikosov and Gorkov (AG)~\cite{abr61} were first to study the effects of randomly distributed paramagnetic impurities on the properties of conventional superconductors. The electron-impurity exchange interaction, $J\,\vec{S}\cdot\vec{s}$, was treated perturbatively in the Born approximation, with $\vec{s}=(\hbar/2)\vec\sigma$ the electron spin operator, $\vec\sigma$ are the Pauli matrics and $\vec{S}$ is the impurity spin.
AG calculated the suppression of the transition temperature $T_c$, and the mean-field superconducting order parameter (``gap'' function), $\Delta(T)$, as a function of the Born scattering rate, $1/\tau_{\text{S}}=\pi N_f J^2 S(S+1)$.
They also showed that for increasing impurity concentration \emph{gapless superconductivity} is possible, i.e. a range of impurity concentrations for which there is a finite density of states down to the Fermi level, yet $T_c$ and $\Delta$ are finite, see also Ref.~\onlinecite{mak69}.
Recent studies connect the appearance of gapless superconductivity to a topological phase transtion~\cite{yer22,yer22a}.

The effects of spin-exchange scattering beyond the Born approximation were pioneered by Yu~\cite{yu65}, Shiba~\cite{shi68} and Rusinov~\cite{rus69a} (YRS). These authors included the exchange interaction to all orders in the amplitude for quasiparticles, including those forming Cooper pairs, scattering off the impurity spin, the latter treated as a classical spin.
For an isolated paramagnetic impurity, exchange scattering generates quasiparticle states with energies below the continuum edge, $\pm\varepsilon_l=\pm\Delta\,\cos(\delta_l^{+}-\delta_l^{-})$, where $\delta_l^{\pm}$ are the phase shifts for quasiparticles with orbital angular momentum $l$ relative to the impurity with spin aligned parallel ($+$) or anti-parallel ($-$) to that of the impurity spin~\cite{rus69a}.
Note that there are impurity states with excitation energy above the Fermi level (particle-like), $0 \le +\varepsilon_l < \Delta$, and impurity states below the Fermi energy, $-\Delta < -\varepsilon_l \le 0$. The negative energy states are occupied for the $T=0$ ground state; thus, hole-like excitations are created at finite temperature and under the action of electromagnetic (EM) fields.
For a finite concentration of randomly distributed impurities the spectrum broadens into an \emph{impurity band}, with bandwidth determined by the impurity concentration, and the center of the band determined primarily by the strength of the exchange interaction~\cite{shi68,mac72}.

\vspace*{-7mm}
\subsection{Outline}
\vspace*{-3mm}

We begin with theoretical methods in Sec.~\ref{sec-Theoreticals_Methods} based on the quasiclassical theory of superconductivity~\cite{eil68,lar69,eli72}, combined with configurational-averaged scattering theory of quasiparticles, including bound pairs, by random distributions of non-magnetic and paramagnetic impurities~\cite{edw58,abr61,he25}. Our formalism, and notation, follows closely that of Ref.~\onlinecite{rai94}.
Section~\ref{sec-Equilibrium_Theory} includes the formulation and development of the T-matrix and self energies for scattering by both non-magnetic and paramagnetic impurities.

These methods are used to calculate the pair-breaking effects of impurities on both equilibrium and nonequilibrium properties of conventional spin-singlet superconductors reported in Secs.~\ref{sec-Tc+Delta} and~\ref{sec-DoS}.
Our focus is on superconducting devices and resonators with \emph{low concentrations} of impurities with quasiparticle-impurity exchange interactions.

In Sec.~\ref{sec-EM_Response} we connect the sub-gap quasiparticle spectrum to the frequency and temperature dependences of the a.c. conductivity for superconducting elements operating at microwave frequencies.
Results for superconducting device performance are based on the causal (retarded) current response to a transverse EM field obtained from the nonequilibrium Keldysh propagator.
The signature of paramagnetic impurities on the penetration of the microwave field into a superconductor is discussed in Sec.~\ref{sec-London_penetration_depth}. In particular, a non-monotonic anomaly in the superfluid fraction at low temperatures is related to the sub-gap spectrum of quasiparticles generated by scattering from paramagnetic impurities.    

Section~\ref{sec-conductivity} includes our results for dissipation at microwave frequencies obtained from the current response of the low-energy band of sub-gap quasiparticles.
In particular, we find that dissipation arises from both intra-band transitions at low temperatures, $\hbar\omega\in\varepsilon_{\text{imp}}\ll\Delta$, where $\varepsilon_{\text{imp}}$ is the impurity band, as well as inter-band transitions between a nearly occupied negative energy band, $-\varepsilon_{\text{imp}}$, and a nearly unoccupied positive energy band, $+\varepsilon_{\text{imp}}$.

In Sec.~\ref{sec-Q_T1_df} we report results for the impact of paramagnetic impurities on the resonance frequency, $\delta f(T)$, and quality factor, $Q(T)$, for high-Q superconducting resonators operating in the GHz range at low-temperatures. 
Quasiparticle scattering by paramagnetic impurities effects the resonant frequency shift of the cavity for temperatures close to $T_c$. The width and depth of the anomalous (negative) frequency shift in Nb-based SRF cavities~\cite{zar23,uek24,baf25} is impacted by exchange scattering from paramagnetic impurities.
We also show that a low-energy sub-gap spectrum of quasiparticles leads to a \emph{residual surface resistance}, $R_s(0)$, and thus a limiting value of $Q$ in good agreement with experimental results on Nb SRF cavities after removal of surface oxides~\cite{rom20}.

Readers not interested in theoretical methods can jump to Sec.~\ref{sec-Tc+Delta}.

\vspace{-5mm}
\section{Theoretical Methods}\label{sec-Theoreticals_Methods}
\vspace{-2mm}

The quasiclassical theory of superconductivity formulated by Eilenberger~\cite{eil68} and Larkin and Ovchinnikov~\cite{lar69} is ideally suited for computational studies of nonequilibrium response of superconductors to electromagnetic fields. The central elements of the theory are the propagators and self energies obtained by quasiclassical reduction of the field theoretical formulation of the theory of superconductivity~\cite{AGD,schrieffer64}, c.f. Ref.~\onlinecite{rai94}. 

The quasiclassical propagators - Matsubara, retarded, advanced and Keldysh ($x=\{M,R,A,K\}$) - are functions of the phase-space variables, $(\vr,\vp_f)$, where $\vr$ is the position in coordiate space and $\vp_f$ is the position \emph{on the Fermi surface}, i.e. the Fermi momentum~\footnote{Hereafter we set $\vp_f\rightarrow\vp$ for momenta on the Fermi surface, and omit the superscript ``M'' for equilibrium propagators and self-energies.},
\be\label{eq-quasiclassical_propagators}
\whmfG^{x}(\vp,\vr;\varepsilon,t)
\equiv
\frac{1}{a}\int\,d\xi_{\vp}\,\tz\,\whG^x(\vp,\vr;\varepsilon,t)
\,,
\ee
where $0<a<1$ is the spectral weight of the normal-state quasiparticle pole centered at excitation energy $\xi_{\vp}=v_f(|\vp|-p_f)$ near the Fermi surface, and
\be
\tx=\begin{pmatrix}0 & \hone \cr \hone & 0 \end{pmatrix}
\,,\,
\ty=\begin{pmatrix}0 & -i\hone \cr i\hone & 0 \end{pmatrix}
\,,\,
\tz=\begin{pmatrix}\hone & 0 \cr 0 & -\hone\end{pmatrix}
\,,
\ee
are the Pauli matrices in particle-hole, ``Nambu'', space and $\hone$ is the unit matrix in spin space~\footnote{$4\times 4$ Nambu matrices are denoted by a widehat, while $2\times 2$ spin matrices are denoted by a narrow hat.}.

Equilibrium states are described by the Matsubara propagator which can be expressed in terms of the quasiparticle and quasihole propgators, the diagonal elements in Nambu space, and the off-diagonal quasiclassical Cooper pair propagators,
\begin{equation}\label{eq-quasiclassical_G} 
\whmfG(\vp,\vr;\varepsilon_n) 
= 
\left(
\begin{array}{cc}
\hat g(\vp,\vr;\varepsilon_n) 
& 
\hat f(\vp,\vr;\varepsilon_n) 
\\
\hat{\bar f}(\vp,\vr;\varepsilon_n) 
& 
\hat{\bar g}(\vp,\vr;\varepsilon_n) 
\end{array}
\right)
\,,
\end{equation}
where $\hat g$ ($\hat f$) is the $2\times 2$ spin matrix quasiparticle (pair) propagator for momenta $\vp$ on the Fermi surface, Matsubara energy, $\varepsilon_n=(2n+1)\pi/\beta$ where $\beta=1/T$ and $T$ is the absolute temperature, as a function of position $\vr$~\footnote{We set $\kb=\hbar=1$ throughout this report. ln (log) denotes the base e (base 10) logarithm.}. The lower components, $\hat{\bar g}$ and $\hat{\bar f}$, define the propagator for hole-like quasiparticles and the conjugate pair propagator, which are related to $\hat g$ and $\hat f$ by Fermion exchange symmetry and particle-hole symmetry~\footnote{See Ref.~\cite{rai94} for a details on these symmetry relations.}.

The equal-time propagator defines the one-particle density matrix in Nambu space, 
\be
\widehat\varrho(\vp,\vr)\equiv\lim_{\tau\rightarrow 0^+}\,
\frac{1}{\beta}\sum_{n}\,\whmfG(\vp,\vr;\varepsilon_n)\,e^{-i\varepsilon_n\tau/\hbar}
\,,
\ee
from which all one-body observables can be calculated. For example, the local electron spin density is given by
\be
\vec{s}(\vr)=N_f\int dS_{\vp}\,\Tr{\widehat\varrho(\vp,\vr)\,\widehat{\vec{s}}}
\,,
\ee
\be
\hspace*{-3mm}
\mbox{where}\quad
\widehat{\vec{s}}\equiv\frac{\hbar}{2}\,\widehat{\vec{\Sigma}} 
=
\frac{\hbar}{2}
\begin{pmatrix}
\vec{\sigma} 
& 0
\cr
0
& \vec{\sigma}^{\text{tr}}
\end{pmatrix}
\,,
\ee
is the quasiparticle spin operator in Nambu space, with $\vec\sigma$ being the $2\times 2$ Pauli spin matrices~\footnote{In the hole-sector $\vec\sigma^{\text{tr}}=-\sigma_y\vec{\sigma}\sigma_y$ is the transpose of the Pauli matrices}, and the integration is an average over the Fermi surface.

Similarly, the local equilibrium Cooper pair amplitude is defined by the equal-time limit of Gorkov's anomalous propagator,
\be
\hat f(\vp,\vr)
\equiv 
\nicefrac{1}{\beta}\lim_{\tau\rightarrow 0^+}\sum_n\,
\hat f(\vp,\vr;\varepsilon_n)\,e^{-i\varepsilon_n\tau/\hbar}
\,.
\ee

Analytic continuation of the diagonal Matsubara propagator to the real energy axis determines the retarded propagator, $\whmfGr(\vp,\vr;\varepsilon)=\whmfG(\vp,\vr;i\varepsilon_n\rightarrow\varepsilon+i0^{+})$, from which the spectral resolution of one-body observables are defined. In particular, the spin-averaged, local density of states for quasiparticles with momenta and excitation energies near the Fermi surface is,
\be
N(\vp,\vr;\varepsilon)
=
-\frac{N_f}{4\pi}\,\Im\,\Tr{\whtauz\whmfGr(\vp,\vr;\varepsilon)}
\,.
\ee 


\vspace*{-5mm}
\subsection{Eilenberger's Equations}
\vspace*{-3mm}

The quasiparticle and anomalous pair propagators, organized into $4\times 4$ Nambu matrices, obey Gorkov's equations~\cite{gor58}. 
Eilenberger transformed Gorkov's equations~\cite{gor58} for the quasiparticle and anomalous pair propagators, into a matrix transport-type equation for the quasiclassical Matsubara propagator defined in Eq.~\eqref{eq-quasiclassical_propagators}~\cite{eil68}, 
\be\label{eq-Eilenberger_Transport}
\hspace*{-3mm}
\commutator{i\varepsilon_n\whtauz\ns-\ns\whSigma(\vp,\vr;\varepsilon_n)}
           {\whmfG(\vp,\vr;\varepsilon_n)}
\ns+\ns 
i\hbar\vv_{\vp}\cdot\grad\whmfG(\vp,\vr;\varepsilon_n)
=0
\,.
\ee
In contrast to Gorkov's equation, which is a second-order differential equation with a unit source term originating from the Fermion anti-commutation relations, Eilenberger's equation is a homogeneous, first-order differential equation describing the evolution of the matrix propagator along classical trajectories defined by the Fermi velocity, $\vv_{\vp}=\grad_{\vp}\xi_{\vp}$.
The physical solutions to the transport equation satisfy Eilenberger's normalization condition~\cite{eil68},
\be\label{eq-Eilenberger_Normalization}
\whmfG(\vp,\vr;\varepsilon_n)^2 = -\pi^2\,\tone
\,,
\ee
which enforces conservation of the spectral weight within the bandwidth of low-energy excitations.  
\vspace*{-3mm}
\subsection{Self Energies}
\vspace*{-3mm}

\begin{figure}[t]
\begin{minipage}{\columnwidth}
\includegraphics[width=\columnwidth]{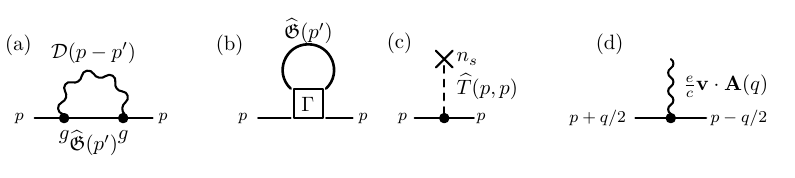}
\vspace*{-5mm}
\caption{Leading self-energy diagrams in the quasiclassical theory of superconductivity. 
(a) electron-phonon self energy,
(b) mean-field electron-electron self energy,
(c) electron-impurity self energy in the T-matrix approximation, and
(d) the coupling of charge currents to an electromagnetic field 
    with $q=(\vq,\omega_m)$.
}
\label{fig-Sigma}
\end{minipage}
\end{figure}

The physical properties of a particular superconducting material are encoded in the self energy, $\whSigma(\vp,\vr;\varepsilon_n)$, that enters Eq.~\eqref{eq-Eilenberger_Transport}. To leading order in the small expansion parameters of quasiclassical theory, $\sml = \{k_{\text{B}} T/E_f,\hbar/p_f\xi_0,\hbar/\tau E_f,\hbar\omega/E_f,\nicefrac{e}{c}v_f A(\omega)/E_f\}$, the processes that define the self-energy are represented by Feynman diagrams for electron-phonon, electron-electron and electron-impurity interactions or the coupling to external fields, are shown in Fig.~\ref{fig-Sigma}.

\vspace{-5mm}
\subsubsection{Pairing Self-Energy}
\vspace{-2mm}

The electron-phonon self-energy, Fig.\ref{fig-Sigma}(a), is central to the theory of superconductivity in elemental superconductors such as Al, Nb, Ta. Phonon-mediated electron-electron interactions in the Cooper channel are often sufficiently attractive, and retarded in time, to overcome the repulsive electron-electron interaction in the Cooper channel, Fig.\ref{fig-Sigma}(b).
For these superconductors the dominant pairing channel is the spin-singlet ($S=0$), orbital $\point{A}{1g}$ channel, i.e. the identity representation.
In what follows we consider pairing mediated by a net attractive interaction in the spin-singlet ($S=0)$, isotropic ``s-wave'' ($L=0$) channel, $g(\varepsilon_n;\varepsilon_n')$, which gives rise to a macroscopic pairing amplitude and mean-field pairing self energy
\be\label{eq-s-wave_gap-equation}
\Delta(\vr;\varepsilon_n)=
T\sum_{n'}\,g(\varepsilon_n-\varepsilon_{n'})\,
\int dS_{\vp'}\,f(\vp',\vr;\varepsilon_{n'})
\,,
\ee
where $g(\omega_m)\equiv\lambda(\omega_m)-\mu^{\star}$, with $\lambda(\omega_m)$ the phonon-mediated electron-electron interaction, and $\mu^{\star}$ the repulsive Coulomb pseudopotential~\cite{rai94}. Note that $\lambda(\omega_m)$ provides a natural cutoff at the maximum phonon energy $\omegaD$.

For Nb, Ta and Al, with $T_c/\omega_D\ll 1$, the effects of retardation are small, and that also holds for the electromagnetic response of Nb at microwave frequencies, $\hbar\omega \ll 2\Delta_0$~\citep{uek25}~\footnote{At much higher frequencies phonons are observable in the optical absorption spectrum for frequencies just above the gap~\citep{lee89,rai94}.}.
Thus, for our purposes we can approximate $g(\varepsilon_n-\varepsilon_{n'})=g\,\Theta(\omegaD-|\varepsilon_n|)\Theta(\omegaD-|\varepsilon_{n'}|)$, in which case we obtain the weak-coupling gap equation for the pairing self energy,
\be
\Delta(\vr)=g\,\,T\ns\ns\sum_{n'}^{|\varepsilon_{n'}|\le\omegaD}
\,\int dS_{\vp'}\,f(\vp',\vr;\varepsilon_{n'})
\,,
\label{eq-weak-coupling_gap-equation}
\ee
where for any self energy functional, the  pair propagator, $f(\vp,\vr;\varepsilon_n)$, is obtained from the solution of Eilenberger's transport equation~\eqref{eq-Eilenberger_Transport} and normalization condition~\eqref{eq-Eilenberger_Normalization}.

\vspace{-5mm}
\subsubsection{Electron-Impurity Self-Energy}
\label{sec-impurity_self-energy}
\vspace{-2mm}

\begin{figure}[t]
\includegraphics[width=\columnwidth]{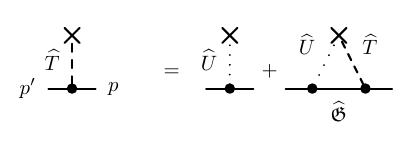}
\caption{T-matrix for multiple scattering of quasiparticles and pairs by an impurity with transition matrix $\whU(\vp',\vp)$. Intermediate states are described by the quasiclassical propagator, $\whmfG(\vp,\varepsilon_n)$.}
\label{fig-Impurity-T-matrix}
\end{figure}

The effects of disorder on superconductivity result from electron scattering off a random distribution of impurities.
The electron-impurity self-energy is determined by the forward scattering limit of the electron-impurity T-matrix and the mean impurity density, $n_s$,
\be\label{eq-Sigma-impurity_T-matrix}
\whSigma_{\text{imp}}(\vp;\varepsilon_n)=n_{s}\widehat{T}(\vp,\vp;\varepsilon_n)
\,,
\ee
where the T-matrix for electron-impurity scattering satisfies a Bethe-Salpeter (BS) equation, shown diagrammatically in Fig.~\ref{fig-Impurity-T-matrix} and given by
\ber
\hspace*{-4mm}
\widehat{T}(\vp',\vp;\varepsilon_n) 
\ns&=&\ns
\widehat{U}(\vp',\vp)
\nonumber\\
\ns&+&\ns 
N_f\ns\int\ns dS_{\vp}''
\widehat{U}(\vp',\vp'')\widehat{\mathfrak{G}}(\vp'',\varepsilon_n)
                       \widehat{T}(\vp'',\vp;\varepsilon_n).
\label{eq-T-matrix}
\eer
This equation encodes the effects of multiple scattering by a \emph{single} impurity, with the intermediate states defined by the self-consistently determined Nambu matrix propagator for both quasiparticles and pairs.

In what follows we consider random distributions of uncorrelated impurities with concentrations of non-magnetic and paramagnetic impurities with mean densities $\nN$ and $\nS$, respectively.
For isotropic impurity potentials the non-magnetic electron-impurity vertex is $V(\vp,\vp')\whone$, while that for spin-exchange scattering is $J(\vp,\vp')\,\vec{S}\cdot\widehat{\vec{s}}$. Both matrix elements are parametrized by scattering amplitudes in orbital angular momentum channels, 
$X(\vp,\vp')=\sum_{l\ge 0}(2l+1)X_l\,P_l(\hat\vp\cdot\hat\vp')\,,\mbox{for}\,\,X=V,J$, where $\{P_l(x)|\,l=0,1,2,\ldots\}$ are Legendre polynomials. 
Amplitudes $V_l$ and $J_l$ define normal-state quasiparticle-impurity scattering phase shifts
$\deltaNl       = \tan^{-1}\bvNl$ where $\bvNl=\pi N_f\,V_l$,
$\deltaSl^{\pm} = \pm\tan^{-1}\buSl$ where 
$\buSl          =\pi N_f J_l\,S\,\frac{\hbar}{2}$.
and $+$ ($-$) corresponds to the electron spin aligned parallel (anti-parallel) to the impurity spin $\vec S$.

The number of relevant channels is determined by the ratio of the atomic radius of the impurity and the Fermi wavelength, i.e. $l\le {\rm Int}\,[k_f R]$, where $k_f=p_f/\hbar$ is the Fermi wavevector~\cite{she16}. Typical values for impurities in metals are $k_f R\approx 1-2$~\cite{mott58}
In what follows we retain only one channel, i.e. the s-wave ($l=0$) channel. We also treat paramagnetic impurities as classical spins. The impurity spins are unordered, in which case we average over the spin orientations of the ensemble of paramagnetic impurities. Thus, $\langle S_i \rangle = 0$, $\langle S_i\,S_j\rangle=S^2\,\nicefrac{1}{3}\delta_{ij}$, etc.
%
\vspace{-5mm}
\subsection{Equilibrium Propagator and Self Energies}
\label{sec-Equilibrium_Theory} 
\vspace{-2mm}

For spatially homogeneous superconducting states, $\grad\whmfG=0$, the propagator satisfies two matrix equations,
\be
\commutator{i\varepsilon_n\whtauz - \whDelta - \whSigma_{\text{imp}}}{\whmfG} =0
\,,\quad\mbox{and}\quad
\whmfG^2 = -\pi^2\,\whone
\,.
\label{eq-homogeneous_Eilenberger}
\ee

Thus, for a spatially homogeneous spin-singlet s-wave superconductor in the \emph{clean limit} ($\whSigma_{\text{imp}}=0$) the pairing self energy is given by $\whDelta=\Delta\,\whtaux\,i\sigma_y$, where the $\gauge$ gauge freedom allows us fix the phase of $\Delta$ to be real, in which case only the $\whtaux$ matrix appears. 
The corresponding solution to Eqs.\eqref{eq-homogeneous_Eilenberger} for the propagator is
\ber
\whmfG(\varepsilon_n) 
&=&
-\pi
\frac{i\varepsilon_n\whtauz - \whDelta}
     {\sqrt{\varepsilon_n^2 + \Delta^2}} 
\,,
\qquad\mbox{(clean limit)}
\label{eq-propagator-clean_limit}
\\
\mbox{where}\quad
\Delta &=& g\,\frac{\pi}{\beta}\,
            \sum_n^{|\varepsilon_n|\le\omegaD}\,
            \frac{\Delta}{\sqrt{\varepsilon_n^2+\Delta^2}}
\,,
\label{eq-gap_equation}
\eer
is the mean-field ``gap function'' in the weak-coupling approximation, $g$ is the pairing interaction in the spin-singlet, s-wave channel and $\omegaD$ is the bandwidth of attraction. 
Standard regularization of the log-divergent sum for $\omega_D/T\rightarrow\infty$ allows us to eliminate the coupling constant, $g$, and the bandwidth $\omega_D$ in favor of the measureable clean limit transition temperature, $T_{c_0}$. The Matsubara sum now converges on the low-energy scale set by $T_{c_o},\Delta\ll\omega_D$,
\be
\ln\left(\frac{T}{T_{c_0}}\right)
=
2\pi T\,\sum_{n=0}^{\infty}
\left[
\frac{1}{\sqrt{\varepsilon_n^2+\Delta^2}}
-
\frac{1}{|\varepsilon_n|}
\right]
\,.
\label{eq-gap_equation2}
\ee

\vspace*{-5mm}
\subsubsection{Non-Magnetic Impurities}
\vspace*{-3mm}

For non-magnetic impurities configurational averaging is defined by a joint probabilty distribution for the random positions of the impurities. For concentrations of \emph{uncorrelated} impurities the configurational averaged T-matrix reduces to the following BS equation for scattering restricted to the s-wave channel,
\be
\TN = \vN\,\whone + N_f\,\vN\,\langle\whmfG(\vp,\varepsilon_n)\rangle\,\TN
\,,
\label{eq-T-matrix_non-magnetic}
\ee
where $\vN$ is the matrix element for electron-impurity scattering by non-magnetic impurities in the s-wave channel, and $\langle\whmfG(\vp,\varepsilon_n)\rangle$ is the Fermi surface average of the propagator, $\whmfG(\vp,\varepsilon_n)$. 
In general the propagator is anisotropic, even for conventional BCS superconductors such as Nb. Non-magnetic impurities are pair-breaking in anisotropic superconductors, including anisotropic BCS superconductors, as described in some detail in Ref.~\cite{zar22}. However, for isotropic (``s-wave''), spin-singlet superconductivity $\langle\whmfG\rangle=\whmfG(\varepsilon_n)$, in which case the T-matrix in Eq.~\eqref{eq-T-matrix_non-magnetic} is easily obtained using the normalization condition in Eq.~\eqref{eq-Eilenberger_Normalization},
\be
\TN = \frac{1}{\pi N_f}\,
\left\{
\frac{\bvN}{1+\bvN^2}\whone
+
\frac{\bvN^2}{1+\bvN^2}\,\whmfG(\varepsilon_n)
\right\}
\,,
\ee
where $\bvN\equiv\pi N_f\vN$. This result is often expressed in terms of the s-wave scattering phase shift $\delta_{\text{N}}=\tan^{-1}\bvN$,
\be
\TN=\frac{1}{\pi N_f}\,
\left\{
\sin\deltaN\cos\deltaN\whone
+
\sin^2\ns\deltaN\,\whmfG(\varepsilon_n)
\right\}
\,.
\ee
The unit matrix term in $\TN$ does not contribute to the impurity self-energy because it drops out of Eilenberger's Eq.~\eqref{eq-homogeneous_Eilenberger}. Thus,
\be
\whSigN=\frac{1}{2\tauN}\whmfG
\,,
\ee
where the quasiparticle-impurity scattering rate is 
\be
\frac{1}{2\tauN} 
=
\frac{\nN}{\pi N_f}\,\frac{\bvN^2}{1+\bvN^2}\equiv\GammaN\,\sin^2\deltaN
\,,
\ee
which in the unitarity limit becomes,
\be
\frac{1}{2\tauN} 
\xrightarrow[|\vN|\rightarrow\infty]{}
\GammaN
\equiv 
\frac{\nN}{\pi N_f}
\,.
\label{eq-nonmag_unitarity-limit}
\ee

\vspace*{-5mm}
\subsubsection{Paramagnetic Impurities}
\label{sec-paramagnetic_impurities}
\vspace*{-3mm}

For paramagnetic impurities the self-energy is defined by an additional configurational average of the T-matrix over the probability distribution of impurity spin directions,
\be
\whSigS(\varepsilon_n)
=
\nS\int\dangle{\vec{S}}\whT(\varepsilon_n;\vec{S})
\equiv 
\nS\,\TS(\varepsilon_n)
\,, 
\label{eq-SigmaS}
\ee
where $\whT(\varepsilon_n;\vec S)$ describes multiple scattering of quasiparticles and Cooper pairs by a single \emph{polarized} magnetic impurity with spin $\vec{S}=S\,\vec{e}$ where $\vec{e}\cdot\vec{e}=e^{i}e^{i}=1$. The ensemble average over a uniform probability distribution for the directions of the impurity spins defines the T-matrix for the unpolarized distribution of \emph{paramagnetic} impurites, $\TS(\varepsilon_n)$.

The symmetry contraints of a homogeneous paramagnetic impurity distribution imply that $\TS(\varepsilon_n)$ contains only terms proportional to $\whtauz$ and $\whtaux\,i\sigma_y$. In particular, (i) there are no terms proportional to $i\whtauy\,i\sigma_y$, as such a term would break time-reversal symmetry by the renormalized gap amplitude, and (ii) there are no spin-triplet terms proportional to $\whtaux\,i\vec\sigma\sigma_y$, which would break spin-rotation symmetry.
Thus, the paramagentic impurity self-energy includes only these terms,
\be\label{eq-Sigma-impurity}
\whSigS(\varepsilon_n) 
=
\SigS(\varepsilon_n)\,\whtauz
+
\DelS(\varepsilon_n)\,\whtaux\,i\sigma_y
\,.
\ee
The quasiclassical propagator then has the same Nambu matrix structure as the clean-limit propagator in the Eq.~\eqref{eq-propagator-clean_limit}, 
\be
\whmfG(\varepsilon_n) =
-\pi
\frac{i\tilde\varepsilon_n\whtauz - \tilde\Delta\whtaux\,i\sigma_y}
     {\sqrt{\tilde\varepsilon_n^2 + \tilde\Delta^2}} 
\,,
\label{eq-propagator-renormalized}
\ee
where the renormalized Matsubara frequency and pairing self energy are given by
\ber
i\tilde\varepsilon_n 
&=&
i\varepsilon_n-\SigN(\varepsilon_n) -\SigS(\varepsilon_n)
\,,
\\
\tilde\Delta
&=&
\Delta+\DelN(\varepsilon_n)+\DelS(\varepsilon_n)
\,. 
\eer

To compute the impurity self energy in Eq.~\eqref{eq-SigmaS} we need to construct a BS equation for the configurational averaged T-matrix, $\TS(\varepsilon_n)$.

\vspace*{-3mm}
\subsubsection{Bethe-Salpeter Equation for $\TS\equiv\langle\whT(\varepsilon_n;\vec{S})\rangle$}
\label{sec-BS_Equation}
\vspace*{-3mm}

\begin{figure}[t]
\includegraphics[width=\columnwidth]{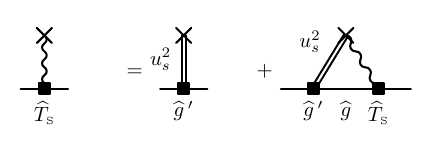}
\caption{Configurational averaged T-matrix for scattering of quasiparticles and pairs from an impurity selected from an ensemble of paramagnetic impurities. 
The leading order term is defined by the vertex $u_s^2\,\whg'$ and 
intermediate states are described by the configurational averaged 
quasiclassical propagator, $\whg=N_f\,\whmfG$.}
\label{fig-PM-Impurity-T-matrix}
\end{figure}

We start from the BS equation for $\whT(\varepsilon_n;\vec S)$ in Eq.~\eqref{eq-T-matrix} for multiple scattering of quasiparticles and Cooper pairs by a magnetic impurity with spin $\vec{S}$. We expand $\whT$ as an infinite series in the spin-exchange interaction matrix element,
\be\label{eq-T-matrix_series}
\whT = \whU 
     + \whU\whg\whU 
     + \whU\whg\whU\whg\whU + 
     + \whU\whg\whU\whg\whU\whg\whU 
     + \ldots 
\,,
\ee
where 
\ber
\whU 
= u_s\,\vec{e}\cdot\widehat{\vec\Sigma}
\,,
\quad
\widehat{\vec\Sigma} = 
\begin{pmatrix}
\vec\sigma & 0 \cr
0 & \vec\sigma^{\text{tr}}
\end{pmatrix}
\,,
\eer
with $u_s\equiv J\,\frac{\hbar}{2}\,S$ and $\whg\equiv N_f\whmfG$.
The equilibrium propagator can be expressed in terms of the quasiparticle and pair propagators,
\ber
\whg
&=&
N_f\whmfG=g_3\whtauz+g_1\whtaux i\sigma_y
\,,
\label{eq-impurity-renormalized_propagator}
\\
g_3
&=&
-\pi\,N_f\,\frac{i\tilde\varepsilon_n}
                {\sqrt{\tilde\varepsilon_n^2+\tilde\Delta^2}}
\,,
\label{eq-g3}
\\
g_1
&=&
+\pi\,N_f\,\frac{\tilde\Delta}
                {\sqrt{\tilde\varepsilon_n^2+\tilde\Delta^2}}
\,,
\label{eq-g1}
\eer
which satisfy the normalization condition, 
\be
\whg^2 = \left(g_3^2 - g_1^2\right)\whone
       = -\pi^2\,N_f^2\,\whone 
\,.
\ee

We carry out the configurational average over the ensemble of paramagnetic impurities term by term in Eq.~\eqref{eq-T-matrix_series}. The first-order term vanishes by symmetry, 
\be
\langle\whU\rangle 
=
\int\dangle{\vec{e}}\,u_s\,\vec{e}\cdot\widehat{\vec\Sigma}
\equiv 0
\,.
\ee
This extends to all terms odd in $\whU\sim\vec{e}\cdot\widehat{\vec\Sigma}$.
Thus, the leading order term for the configurational averaged T-matrix is 2$^{\text{nd}}$ order in $u_s$,
\be
\langle\whU\whg\whU\rangle 
=u_s^2\int\dangle{\vec{e}}\,e^i\,e^j\,\whSigma^i\whg\whSigma^j
=u_s^2\,\nicefrac{1}{3}\,\whSigma^i\,\whg\,\whSigma^i
\equiv
u_s^2\,\whg'
\,.
\ee
It is straightforward to show that 
\be
\whg'=N_f\whmfG'= g_3\whtauz - g_1\whtaux i\sigma_y
\,.
\label{eq-spin-flip_propagator}
\ee
Comparing this result with Eq.~\eqref{eq-impurity-renormalized_propagator} we see that the 2$^{\text{nd}}$ order Born scattering term leads to a sign change in the off-diagonal spin-singlet pairing self energy, and is responsible for pair breaking from this double spin-exchange process.
Note that $\whg'$ also satisfies the normalization condition, $(\whg')^2=-\pi^2 N_f^2\,\whone$.

Similarly, the 4$^{\text{th}}$ order contribution to the configurational averaged T-matrix reduces to,
\be
\langle \whU\whg\whU\whg\whU\whg\whU \rangle 
=
u_s^2\whg'\,\times \whg\,\times u_s^2\whg'
\,,
\ee
and similarly for all even powers of $\whU$. The resulting configurational averaged T-matrix satisfies the BS equation, 
\be
\TS = u_s^2\,\whg' + u_s^2\,\whg'\,\times \whg\,\times \TS
\,,
\label{eq-T-matrix_configurational-averaged}
\ee
shown diagramatically in Fig.~\ref{fig-PM-Impurity-T-matrix}. The formal solution for $\TS$ is then,
\ber
\TS = \left(\whone-u_s^2\,\whg'\,\whg\right)^{-1}\times u_s^2\,\whg' 
\,.
\eer
The matrix inverse $\left(\whone - u_s^2\,\whg'\,\whg\right)^{-1}$ is easily obtained by making use of the normalization conditions, $(\whg)^2 = (\whg')^2 = -\pi^2\,N_f^2\,\whone$, and the anti-commutator,
\be
\hspace*{-3mm}
\anticommutator{\whg}{\whg'} 
= 2\,\left(g_3^2 - g_1^2\right)\,
\whone 
= 2\pi^2\,N_f^2\,
\left\{1-2\frac{\tilde\varepsilon_n^2}
	       {\tilde\varepsilon_n^2+\tilde\Delta^2}
\right\}
\whone 
\,,
\ee
where the second equality follows from Eqs.~\eqref{eq-g3}-\eqref{eq-g1}. 
Thus, it is straight-forward to construct the inverse,
\ber
\left(\whone-u_s^2\,\whg'\,\whg\right)^{-1}
=
\frac{\left(\whone-u_s^2\,\whg\,\whg'\right)}
{(1-\bar{u}_s^2)^2 
      + 4\,\bar{u}_s^2\,
\left(\displaystyle{\frac{\tilde\varepsilon_n^2}
	   {\tilde\varepsilon_n^2+\tilde\Delta^2}}\right)}
\,,
\eer
where $\bar{u}_s\equiv\pi N_f\,u_s$. We then obtain the configurational averaged T-matrix, 
\ber
\TS
&=& u_s^2 
\frac{\bar{u}_s^2}
     {(1-\bar{u}_s^2)^2 + 4\,\bar{u}_s^2\,
      \left(\displaystyle{\frac{\tilde\varepsilon_n^2}
	   {\tilde\varepsilon_n^2+\tilde\Delta^2}}\right)}\times\whg
\nonumber\\
&+& u_s^2 
\frac{1}
     {(1-\bar{u}_s^2)^2 + 4\,\bar{u}_s^2\,
      \left(\displaystyle{\frac{\tilde\varepsilon_n^2}
	   {\tilde\varepsilon_n^2+\tilde\Delta^2}}\right)}\times\whg'
\,.
\label{eq-T_S-matrix}
\eer
This result when multiplied by the paramagnetic impurity density, $n_s$, and analytically continuted to the real energies: $i\varepsilon_n\rightarrow\varepsilon+i0^+$, is equivalent to Shiba's Eq.~(3.6)~\cite{shi68} for the self energy.
The poles and branch cuts of $\TS^{\text{R}}(\varepsilon)=\TS(\varepsilon_n\rightarrow -i\varepsilon+0^+)$ reveal the spectrum of both continuum excitations and sub-gap bound states.
In particular, in the limit, $\nS\rightarrow 0$ (isolated impurities), we can neglect the impurity renomalization of $\varepsilon$ and $\Delta$, in which case the denominator of $\TS^{\text{R}}(\varepsilon)$ has branch cuts at continuum edges, $\pm\Delta$, and isolated poles at sub-gap energies,
\be
\varepsilon^{\pm} = 
\pm \Delta\,\frac{1-\bar{u}_s^2}{1+\bar{u}_s^2}
\,,
\label{eq-bound-state-poles}
\ee
corresponding to quasiparticle states bound to the impurity with wave functions of radial extent of order $\xi_{\text{S}}= \xi_0/|\sin(\delta^+-\delta^-)|$, where $\xi_0=\hbar v_f/\pi\Delta_0$ is the zero-temperature coherence length in the clean limit, and the phase shifts for pure exchange scattering are defined by $\tan\delta^{\pm} = \pm\bar{u}_s$. Thus, for $\bar{u}_s=\pm 1$ there is a bound state at the Fermi energy with radial extent $\xi_0$. For $0< |\bar{u}_s| < 1$ the bound state energy lies in the gap and the wave function spreads radially.
This result is equivalent to that of Rusinov~\cite{rus69a} for impurities with pure exchange interactions with quasiparticles.
Our result for the T-matrix for multiple scattering by a paramagnetic impurity that includes scattering of 
conduction electrons in both spin-flip (exchange) and non-spin-flip (potential) channels has the same 
structure as that in Eq.~\eqref{eq-T_S-matrix}, agrees with Rusinov's result in the single impurity 
limit, and is discussed in App.~\ref{app-two-channel_T-matrix}.

For a finite concentration of paramagnetic impurities the bound states hybridize, generating an inhomogeneous band of quasiparticle states centered near the single-impurity bound state energy.
The bandwidth and spectral weight of these sub-gap states is encoded in the renormalization of $\varepsilon$ and $\Delta$ by the impurity self energy, 
\be
\whSigS \equiv \nS\,\TS(\varepsilon_n)
=
\SigS\whtauz+\DelS\whtaux\,i\sigma_y
\,.
\ee
Note that the first term in Eq.~\eqref{eq-T_S-matrix}, proportional to $\whg$, is a correction to the potential scattering T-matrix from the exchange interaction. For isotropic, spin-singlet superconductors this term is non-pair-breaking and drops out of the equilibrium Eilenberger equation~\eqref{eq-Eilenberger_Transport}, and thus does not contribute to pair breaking and the sub-gap spectrum. 
%
\begin{figure}[t!]
\includegraphics[width=\columnwidth]{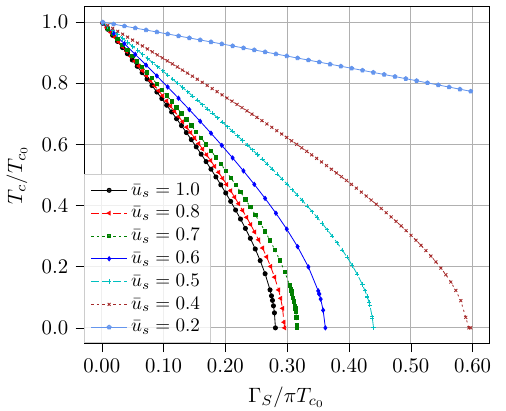}
\caption{Suppression of the transition temperature, $T_c$, as a function of $\GammaS$ (impurity concentration) for a range of $\buS$ spanning the Born limit, $\buS\ll 1$, to maximal pair-breaking for $\buS=1$.}
\label{fig-Tc_vs_GammaS}
\end{figure}
%
However, this exchange correction, as well as the non-magnetic impurity scattering self energy, does contribute to the non-equilibrium response functions, and thus to the microwave conductivity as we discuss in Sec.~\ref{sec-EM_Response}. Thus, we retain both sets of terms in the self energy obtained from the configurational averaged 
T-matrix, 
\ber
\hspace*{-5mm}
\Sigma_{\text{imp}} 
\equiv
\SigN+\SigS
&=&
-\frac{1}{2\tau_1}\,
\frac{i\tilde\varepsilon_n}{\sqrt{\tilde\varepsilon_n^2 +\tilde\Delta^2}}
\,,
\label{eq-Sigma_imp-tau1}
\\
\hspace*{-5mm}
\Delta_{\text{imp}}
\equiv
\DelN+\DelS
&=&
+\frac{1}{2\tau_2}\,
\frac{\tilde\Delta}{\sqrt{\tilde\varepsilon_n^2 +\tilde\Delta^2}}
\,,
\label{eq-Delta_imp-tau2}
\eer 
where the scattering rates are given by 
\ber
\hspace*{-5mm}
\frac{1}{\tau_1} 
&=& 
 \frac{1}{\tauN}
+\frac{1}{\tauS}
\frac{1+\buS^2}
     {(1+\buS^2)^2 - 4\,\buS^2\,
      \left(\displaystyle{\frac{\tilde\Delta^2}
	   {\tilde\varepsilon_n^2+\tilde\Delta^2}}\right)}
\,,
\label{eq-tau1}
\\
\hspace*{-5mm}
\frac{1}{\tau_2} 
&=&
 \frac{1}{\tauN}
-
\frac{1}{\tauS}
\frac{1-\buS^2}
     {(1+\buS^2)^2 - 4\,\buS^2\,
      \left(\displaystyle{\frac{\tilde\Delta^2}
	   {\tilde\varepsilon_n^2+\tilde\Delta^2}}\right)}
\,,
\label{eq-tau2}
\eer

\vspace*{-5mm}
\be
\hspace*{-16mm}
\mbox{with}
\;\;
\frac{1}{2\tauN}
=
\frac{\nN}{\pi N_f}\frac{\bvN^2}{1+\bvN^2}
\;\;\:
\mbox{and}
\;\;\;
\frac{1}{2\tauS}
=
\frac{\nS}{\pi N_f}\,\buS^2
\,,
\label{eq-tauN-tauS}
\ee
are the normal-state scattering rate for non-magnetic impurities with density $\nN$, and the Born scattering rate for paramagnetic impurities with density $\nS$.
Equations~\eqref{eq-tau1} and \eqref{eq-tau2} are not restricted to the Born approximation. The dimensionless matrix elements, $\bvN=\pi N_f\vN$ and $\buS=\pi N_f\uS$ are measures of the strength of the potential and exchange interaction between quasiparticles and impurities.
Equations~\eqref{eq-Sigma_imp-tau1},~\eqref{eq-Delta_imp-tau2}, when analytically continued to the real energy axis, $i\varepsilon_n\rightarrow\varepsilon+i0^+$, agree with Shiba's Eqs.~(3.6) and (3.7) for pure exchange scattering by a distriburion of paramagnetic impurities.
And if we retain only terms of order $\bar{u}_s^2$ in Eqs.~\eqref{eq-Sigma_imp-tau1} and~\eqref{eq-Delta_imp-tau2} we obtain the Born limit studied by Abrikosov and Gorkov (AG).
Thus, the renormalized Matsubara energy and pairing self energy become,
\ber
\tilde\varepsilon_n
&=&
\varepsilon_n 
+ 
\frac{1}{2\tau_1}
\frac{\tilde\varepsilon_n}{\sqrt{\tilde\varepsilon_n^2+\tilde\Delta^2}}
\,,
\label{eq-renormalized_epsilon}
\\
\tilde\Delta
&=&
\Delta
+
\frac{1}{2\tau_2}
\frac{\tilde\Delta}{\sqrt{\tilde\varepsilon_n^2+\tilde\Delta^2}}
\,.
\label{eq-renormalized_Delta}
\eer

We introduce renormalization factors: $\tilde\varepsilon_n\equiv Z_3(\varepsilon_n)\,\varepsilon_n$ and $\tilde\Delta\equiv Z_1(\varepsilon_n)\,\Delta$, and obtain coupled transcendental equations for $Z_{1,3}$ to be solved as functions of $\varepsilon_n$, $\Delta$, $1/\tau_{\text{N}}$ and $1/\tau_{\text{S}}$. 
The difference in scattering rates in Eqs.~\eqref{eq-renormalized_epsilon} and~\eqref{eq-renormalized_Delta} implies $Z_3\ne Z_1$, and the violation of Anderson's theorem, the suppression of $T_c$, pair-breaking and the existence of quasiparticle states in the gap.

Note that the renormalized scattering rates, $1/\tau_{1,2}$, depend only on the ratio $Z=Z_3/Z_1$, in which case Eqs.~\eqref{eq-tau1} and \eqref{eq-tau2} become,
\ber
\hspace*{-5mm}
\frac{1}{\tau_1} 
&=& 
 \frac{1}{\tauN} 
+\frac{1}{\tauS} 
\frac{1+\buS^2}
     {(1+\buS^2)^2 - 4\,\buS^2\,
      \left(\displaystyle{\frac{\Delta^2}
	   {\bar\varepsilon_n^2+\Delta^2}}\right)}
\,,
\label{eq-tau1_Z}
\\
\hspace*{-5mm}
\frac{1}{\tau_2} 
&=&
 \frac{1}{\tauN} 
-\frac{1}{\tauS} 
\frac{1-\buS^2}
     {(1+\buS^2)^2 - 4\,\buS^2\,
      \left(\displaystyle{\frac{\Delta^2}
	   {\bar\varepsilon_n^2+\Delta^2}}\right)}
\,,
\label{eq-tau2_Z}
\eer
where 
\vspace*{-8mm}
\be
\frac{\tilde\varepsilon_n}{\tilde\Delta} 
=
\frac{Z_3}{Z_1}\frac{\varepsilon_n}{\Delta}
\equiv
\frac{\bar\varepsilon_n}{\Delta}
\,.
\ee 

We obtain $\bar\varepsilon_n$ from the ratio of Eq.~\eqref{eq-renormalized_epsilon} and 
Eq.~\eqref{eq-renormalized_Delta},
\ber
\bar\varepsilon_n
=
\varepsilon_n 
\ns+\ns 
\frac{1}{2}\left(\frac{1}{\ns\tau_1}\ns-\ns\frac{1}{\tau_2}\ns\right)
\frac{\bar\varepsilon_n}{\sqrt{\bar\varepsilon_n^2+\Delta^2}}
\ns\equiv
\varepsilon_n 
+ 
\frac{1}{\bar\tauS}\,
\frac{\bar\varepsilon_n}{\sqrt{\bar\varepsilon_n^2+\Delta^2}}
\,,
\label{eq-bar_epsilon}
\\
\mbox{where}
\quad
\frac{1}{\bar\tau_{\text{S}}}
\equiv
\frac{1}{\tauS} 
\frac{1}{(1+\buS^2)^2 
        -4\,\buS^2\,\left(\displaystyle{\frac{\Delta^2}
	                               {\bar\varepsilon_n^2+\Delta^2}}\right)}
\,.
\label{eq-bar_tauS}
\eer
There are several observations to be made.
(i) A key result is that \emph{if} $1/\tau_1=1/\tau_2$ impurity 
      scattering drops out of all equilibrium properties. This is
      Anderson's theorem.
(ii) The flip side is that \emph{only} the difference between the 
      scattering rates $1/\tau_1$ and $1/\tau_2$, which is proportional 
      to the exchange scattering rate $1/\tauS$, contributes 
      to pair breaking, suppression of $T_c$ and the generation 
      of a spectrum of sub-gap quasiparticles.
(iii) Eq.~\eqref{eq-bar_epsilon} reduces two self-consistency 
      equations for $\tilde\varepsilon_n$ and $\tilde\Delta$ to 
      one self-consistency equation for $\bar\varepsilon_n$. 
Note also that Eqs.~\eqref{eq-bar_epsilon} and~\eqref{eq-tauN-tauS} reduce to 
\be
\bar\varepsilon_n
=
\varepsilon_n 
+
\frac{1}{\tauS}\,
\frac{\bar\varepsilon_n\,\sqrt{\bar\varepsilon_n^2+\Delta^2}}
     {(1+\buS^2)^2\bar\varepsilon_n^2+(1-\buS^2)^2\Delta^2}
\,.
\label{eq-bar_epsilon2}
\ee
Analytic continuation to the real axis, $i\varepsilon_n\rightarrow \varepsilon+i0^+$, highlights the presence of the band of sub-gap states, which in the limit $\nS\rightarrow 0^+$ correspond to the isolated impurity bound states above and below the Fermi level, Eq.~\eqref{eq-bound-state-poles}, found by Rusinov.

\vspace*{-5mm}
\subsubsection{Suppression of $T_c$ and the Order Parameter $\Delta$}
\label{sec-Tc+Delta}
\vspace*{-3mm}

The equilibrium solution to the homogeneous Eilenberger equations including scattering by non-magnetic and paramagnetic impurities reduces to   
\ber
\whmfG(\varepsilon_n) 
&=&
-\pi
\frac{i\bar\varepsilon_n\whtauz - \Delta\whtaux i\sigma_y}
     {\sqrt{\bar\varepsilon_n^2 + \Delta^2}} 
\,,
\label{eq-propagator+impurity_scattering}
\\
\mbox{where}\quad
\Delta &=& \lambda\,\frac{\pi}{\beta}\,
            \sum_n^{|\varepsilon_n|\le\omegaD}\,
            \frac{\Delta}{\sqrt{\bar\varepsilon_n^2+\Delta^2}}
\,,
\label{eq-gap_equation+impurity_scattering}
\eer
The gap equation now includes the pair-breaking impurity self energy with $\bar\varepsilon_n$ defined by Eq.~\eqref{eq-bar_epsilon2}. 
As was done for the gap equation in the clean limit we eliminate the coupling constant and bandwidth in place of $T_{c_0}$,
\be
\ln\left(\frac{T}{T_{c_0}}\right)
=
2\pi T\,\sum_{n=0}^{\infty}
\left[
\frac{1}{\sqrt{\bar\varepsilon_n^2+\Delta^2}}
-
\frac{1}{|\varepsilon_n|}
\right]
\,.
\label{eq-gap_equation+impurity_scattering2}
\ee
This equation for $\Delta$ and Eq.~\eqref{eq-bar_epsilon2} for $\bar\varepsilon_n$ must be solved together self-consistently as a function of $T/T_{c_0}$, the scattering rate $1/\tauS$, and the exchange interaction $\buS$.
The gap parameter $\Delta\rightarrow\,0^+$ as $T\rightarrow T_c$ from below. In this limit $\varepsilon_n=2\pi T_c (n+\nicefrac{1}{2})$ and
\be
\bar\varepsilon_n = \sgn(\varepsilon_n)
                    \left(|\varepsilon_n|+\frac{1}{\tauSn}\right)
\,,
\ee
\be
\mbox{where}\quad
\frac{1}{\tauSn} 
= 
\frac{1}{\tauS}\,\frac{1}{(1+\buS^2)^2}
= 
\GammaS\,\frac{4\buS^2}{(1+\buS^2)^2}
\,,
\label{eq-exchange_scattering_rate}
\ee
is the normal-state quasiparticle-impurity exchange scattering rate, where
\be
\GammaS\equiv\frac{\nS}{2\pi N_f}
\,,
\ee
depends on the concentration of paramagnetic impurities and normal-state quasiparticle density of states at the Fermi level, and plays the same role as the unitarity limit for s-wave, non-magnetic impurity scattering in unconventional superconductors~\cite{xu95}.
In particular, the paramagnetic scattering rate is maximized with respect to the exchange interaction at $\buS^2=1$, and thus $\GammaS$ is the maximum scattering rate for fixed concentration.
Note also that the scattering rate for $\buS^2>1$ is equivalent in terms of equilibrium pair-breaking to the scattering rate for an exchange interaction $1/\buS^2<1$.

\begin{figure}[t!]
\includegraphics[width=\columnwidth]{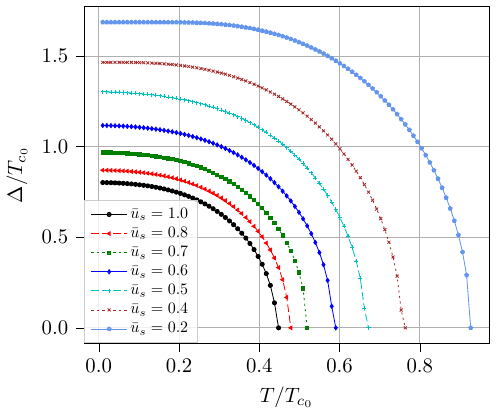}
\caption{Pair-breaking suppression of the superconducting gap $\Delta$ by paramagnetic impurities for $\GammaS/\pi T_{c_0}=0.2$ and the same range of $\buS$ as in Fig.~\ref{fig-Tc_vs_GammaS}.}
\label{fig-Delta_vs_T}
\end{figure}

The physical signficance of $\buS^2=1$ was pointed not long after the works of YRS.
For a single paramagnetic impurity in a spin-singlet, s-wave superconductor $\buS=1$ corresponds to level crossing at the Fermi level, and was interpreted as a quantum phase transition of the ground state by Sakurai~\cite{sak70}.
Salkola et al.~\cite{sal97} carried out a self-consistent analysis for a single impurity spin, including gap suppression near the impurity, and found that the critical point is suppressed, ${\buS}^{\text{critical}}\lesssim 1$.
Our self-consistent analysis for dilute concentrations of paramagnetic impurities also shows suppression of the critical point. In this case the transition occurs 
when the low-energy edge of the positive energy impurity band crosses the Fermi level.

The resulting equation for $T_c$ reduces to
\be
\ln\left(T_{c_0}/T_c\right) 
= 
\psi(\nicefrac{1}{2}+\nicefrac{1}{2\pi\tauSn T_c})
 -
\psi(\nicefrac{1}{2})
\,,
\label{eq-Tc_AG}
\ee
where $\psi(z)$ is the diGamma function~\cite{abramowitz72}.
Equation~\eqref{eq-Tc_AG} agrees with the AG result~\cite{abr61} in the Born limit, $\tauSn\rightarrow\tauS$. The result for the suppression of $T_c$ depends only on the ratio of the exchange scattering rate, $1/\tauSn$, to the pair formation rate, $1/\tau_0=2\pi T_{c_0}$.
For small scattering rates, $1/2\pi\tauSn T_{c_0} \ll 1$, the suppression of $T_c$ is a small and reduces to, 
\be
\frac{T_c}{T_{c_0}}
\simeq 
1-\frac{\pi^2}{4}\,\frac{1}{\pi\tauSn T_{c_0}} 
=
1-\frac{\pi^2}{4}\,\frac{\GammaS}{\pi T_{c_0}}\,
                   \frac{4\buS^2}{(1+\buS^2)^2} 
\,.
\ee
For sufficiently large exchange scattering rates superconductivity is destroyed at a critical value of the scattering rate $1/\pi\tauSn T_{c_0}\vert_{\text{crit}}=\nicefrac{1}{2}e^{-\gamma_{\text{E}}}\approx 0.281$.
Although $T_c/T_{c_0}$ is a function of $1/\pi\tauSn T_{c_0}$, properties of the superconducting state, including $T_c$, the gap amplitude, density of states, conductivity etc. generally depend on both the concentration of impurities, $\nS$, and the exchange interaction, $\buS$. 

The suppression of $T_c$ by quasiparticle scattering from paramagnetic impurities is shown in Fig.~\ref{fig-Tc_vs_GammaS} as a function of $\GammaS\propto\nS$ and exchange interaction, $\buS$, the latter ranging from the Born limit to maximal pair-breaking. For fixed impurity concentration the maximum suppression of $T_c$ occurs for $\buS^2=1$.
Similarly, the mean-field order parameter (gap amplitude $\Delta$) defined by the self-consistent solution of Eqs.~\eqref{eq-gap_equation+impurity_scattering2} and~\eqref{eq-bar_epsilon2} is also suppressed relative to the clean limit gap amplitude at all temperatures as show in Fig.~\ref{fig-Delta_vs_T}. 

\begin{figure}[t!]
\includegraphics[width=\columnwidth]{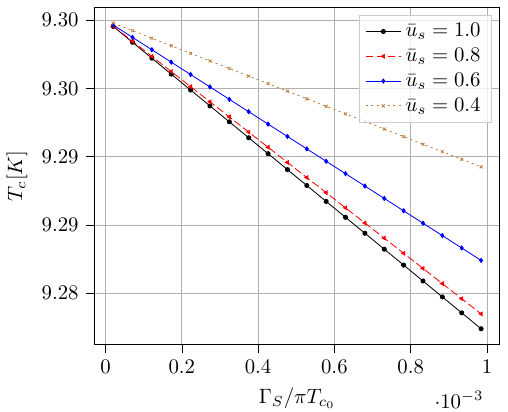}
\caption{Weak suppression of $T_c$ for niobium ($T_{c_0}=9.3\,\mbox{K}$) for low concentrations of paramagnetic impurities, $\GammaS/\pi T_{c_0}\le 10^{-3}$, and $\buS$ as shown in the legend.}
\label{fig-Tc_vs_GammaS-Nb}
\end{figure}

Figure~\ref{fig-Tc_vs_GammaS-Nb} shows the suppression of the transition temperature of niobium for low concentrations of paramagnetic impurities.
The small suppression of $T_c$, even for $\buS^2=1$, is two orders of magnitude smaller than the suppression of $T_c$ by non-magnetic impurity scattering and gap ansiotropy in high-Q superconducting Nb cavities as shown in Fig. 12 of Ref.~\cite{zar22}.
However, as we discuss in Sec.~\ref{sec-conductivity}, in contrast to non-magnetic impurity scattering, very low concentrations of paramagnetic impurities with intermediate exchange coupling can have a major impact on the quality factor and performance of superconducting Nb resonators and devices operating at GHz frequenciesand low temperatures.

\vspace*{-5mm}
\subsubsection{Quasiparticle density of states}
\label{sec-DoS}
\vspace*{-3mm}

The suppression of $T_c$ and $\Delta$ is a reflection of the breaking of Cooper pairs by the exchange interaction between the quasiparticle spin and that of the paramagnetic impurities, and results in the generation of un-paired quasiparticle states with sub-gap energies, $-\Delta < \varepsilon < +\Delta$.

The quasiparticle density of states is obtained from the retarded propagator by analytic continuation of the Matsubara propagator in Eq.~\eqref{eq-propagator+impurity_scattering} to the real axis,
\ber
N(\varepsilon) 
&=& 
N_f\,\Im{\frac{\bepsR(\varepsilon)}{\sqrt{\Delta^2-\bepsR(\varepsilon)^2}}}
\,,
\\
\bepsR(\varepsilon) 
&=& 
\varepsilon 
+
4\GammaS\,\bar\uS^2\,
\frac{\bepsR(\varepsilon)\sqrt{\Delta^2 - \bepsR(\varepsilon)^2 }}
{(1-\buS^2)^2\Delta^2 - (1+\buS^2)^2\bepsR(\varepsilon)^2}
\,.
\label{eq-bar_epsilonR}
\eer

The exchange interaction, $\buS$, generates a pair of impurity bands - one above and one below the Fermi level - centered about the energy of an isolated impurity level (Eq.~\eqref{eq-bound-state-poles}) with bandwidth determined by the impurity density via $\GammaS=\nS/2\pi N_f$.

\begin{figure}[t!]
\includegraphics{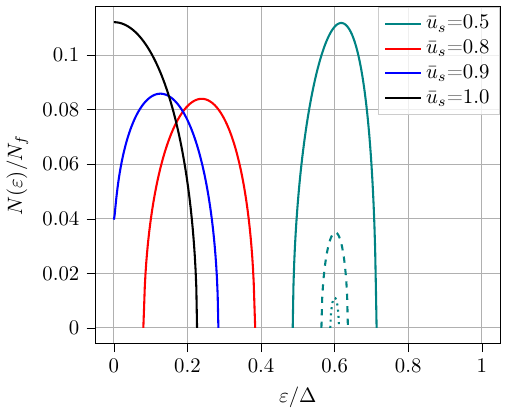}
\caption{Density of states for $T/T_{c}=0.02$ and $\GammaS/2\pi T_{c_0}=3.5\times 10^{-3}$ (solid lines) shows the shift of the sub-gap band toward the Fermi energy for $\buS$ increasing from $0.5\rightarrow 1.0$. The dashed (dotted) line corresponding to $\GammaS/2\pi T_{c_0}=3.5\times 10^{-4}$ ($3.5\times 10^{-5}$) shows the reduction in bandwidth and spectral weight for $\buS=0.5$.}
\label{fig-subgap_spectrum_vs_uS}
\end{figure}

In the Born limit, $\buS^2 \ll 1$, the sub-gap spectrum builds from continuum edges at $\pm\Delta$~\cite{amb65}. With increasing $|\buS|$ the impurity band separates from the continuum and shifts toward the Fermi level as shown in Fig.~\ref{fig-subgap_spectrum_vs_uS}.
In the limit $\buS^2=1$, i.e. maximum pair-breaking at fixed $\nS$, the two bands merge to a single band centered at the Fermi energy defined by the solution of
\be
\bepsR(\varepsilon) 
= 
\varepsilon 
-
\GammaS\,
\frac{\sqrt{\Delta^2 - \bepsR(\varepsilon)^2}}
{\bepsR(\varepsilon)}
\,.
\label{eq-bar_epsilonR_uS=1}
\ee

\begin{figure}[t!]
\includegraphics{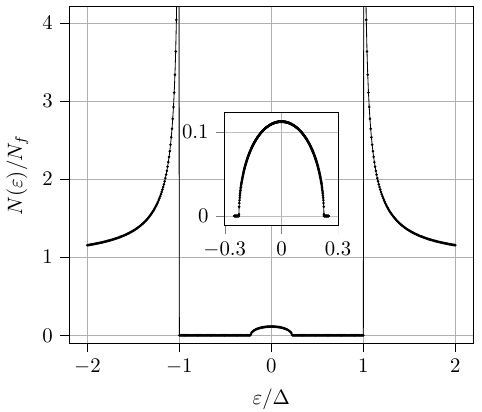}
\caption{Quasiparticle density of states for $T/T_{c}=0.02$ and exchange coupling, $\buS^2=1$. The half-width of the sub-gap impurity band, $\gamma$, corresponds to $\GammaS/2\pi T_{c_0}=3.51\times 10^{-3}$.}
\label{eq-DOS_sub-gap_epsilon=0}.
\end{figure}

The result is a spectrum of gapless quasiparticle states centered at $\varepsilon=0$ as shown in Fig.~\eqref{eq-DOS_sub-gap_epsilon=0}.
The finite DOS at the Fermi level leads to electronic contributions to the heat capacity~\cite{ali11} and thermal conductivity~\cite{gra96a} that are linear in $T$ at low temperatures, $T\ll\gamma$, where $\gamma$ is the half width of the impurity band.
The latter is defined by the imaginary part of $\bepsR(\varepsilon)=\varepsilon+i\gamma$, which for the band centered at $\varepsilon=0$ is
\ber
\gamma \simeq \sqrt{\GammaS\,\Delta}\,
       \ll\Delta\,,\quad\mbox{for}\quad\GammaS\ll\Delta
\,.
\label{eq-gamma-uS=1}
\eer
The number of sub-gap states is obtained by integration over the sub-gap energy band,
\be
N_{\text{bs}}(\varepsilon) = N(0)\,\sqrt{1-(\varepsilon/\varepsilon_{\text{bs}})^2}\,\Theta(\varepsilon_{\text{bs}}-|\varepsilon|)
\,,
\ee
with $N(0)=N_f\,(\varepsilon_{\text{bs}}/\Delta)$ and $\varepsilon_{\text{bs}}=2\gamma$, in which case
\be
\int_{0}^{\varepsilon_{\text{bs}}}\ns\ns\,d\varepsilon\,N_{\text{bs}}(\varepsilon)=\nS 
\,,
\ee
yielding a pair of quasiparticle states per impurity.

For relatively low impurity concentrations a narrow sub-gap band of states shifts away from the Fermi level as the exchange interaction is reduced. The shift in the  position of the impurity band is shown in Fig.~\eqref{fig-subgap_spectrum_vs_uS} for ($\buS=0.5$) to ($\buS=1$) exchange interactions and fixed concentration, $\GammaS/2\pi T_{c_0}=3.5\,\times 10^{-3}$. Even for intermediate exchange interactions, $\buS^2\approx 1$, low concentrations of paramagnetic impurities do not significantly suppress $T_c$ and the gap amplitude, $\Delta$. However, the low-energy sub-gap states can have a dramatic effects on the conductivity at low temperatures and GHz frequencies, and thus the performance of superconducting devices as discussed in Secs.~\ref{sec-conductivity} and~\ref{sec-Q_T1_df}.

\vspace*{-3mm}
\section{Electromagnetic Response}\label{sec-EM_Response}
\vspace*{-3mm}

The quasiclassical transport equations can be extended to calculate the response functions of a superconductor subject to an electromagnetic field (EM). Here we restrict the analysis to transverse EM fields. The interaction potential is then given by the coupling of the vector potential to the charge current,
\begin{equation}
\whSigma_{\text{EM}}(\vr,t)=-\frac{e}{c}\vv_{\vp}\cdot\vA(\vr,t)\whtauz
\,, 
\label{eq-vEM}
\end{equation}
where $e\vv_{\vp}\whtauz$ is the Nambu charge current operator~\footnote{The sign convention used here is: $e=-|e|$ is the charge of the electron} and $\vA(\vr,t)$ satisfies transverse condition, $\grad\cdot\vA=0$.

The quasiclassical propagator is extended to a function of two frequencies, $\whmfG(\vp,\vr;\varepsilon_n)\rightarrow \whmfG(\vp,\vq;\varepsilon_n,\omega_m)$, where $\omega_m$ is the Bosonic Matsubara frequency associated with Fourier mode $\vq$ of the external EM field, $\vA(\vq,\omega_m)$. 
Analytic continuation of $\whmfG(\varepsilon_n;\omega_m)$ to real energies: 
$i\varepsilon_n\rightarrow\varepsilon+i0^+$ with $i\omega_m\rightarrow\omega+i0^+$ yields the Keldysh propagator~\cite{sau17},
\be
\frac{1}{\beta}\sum_{\varepsilon_n}\,\whmfG(\varepsilon_n,\omega_m)
\xrightarrow[i\omega_m\rightarrow\omega+i0^{+}]{}
\int_{-\infty}^{+\infty}\,\frac{d\varepsilon}{4\pi i}\,
\whmfG^K(\varepsilon;\omega)
\,.
\label{eq-analytic_continuation}
\ee

\begin{figure}
\includegraphics[width=\columnwidth]{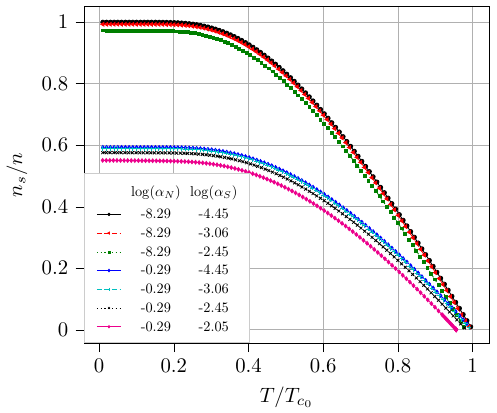}
\caption{The superfluid fraction, $n_s/n$, as a function of $T/T_c$ for frequency $\omega=0.005\,T_{c_0}$. Results are shown for low- and high-concentrations of non-magnetic impurities in the unitarity limit 
[Eq.~\eqref{eq-nonmag_unitarity-limit}] with $\alphaN=\GammaN/\pi T_{c_0}$, and a range of low 
concentrations, $\alphaS=\Gamma_S/2\pi T_{c_0}$, of paramagnetic impurities in the limit $\buS=1$.}
\label{fig-ns_vs_T}
\end{figure}

From $\whmfG^K$ we obtain the retarded (causal) current response as a function of wavevector $\vq$ and frequency $\omega$ of the EM field,
%
\be
\hspace*{-2mm}
\vJ^R(\vq,\omega)
\ns=\ns 2N_f
\ns\int\ns dS_{\vp}
\ns\int\ns\ns\frac{d\varepsilon}{4\pi i}
\nicefrac{1}{4}\Tr{e\vv_{\vp}\whtauz\whmfG^{K}(\vp,\vq;\varepsilon,\omega)}
\,.
\label{eq-Keldysh_Current_Response}
\ee

The trace is over the charge current operator in Nambu space and the integration is over the Fermi surface and all energies. The Keldysh function encodes both the spectral density of the charge carriers and the occuption of these states, in general for nonequilibrium distributions.

The general theory for the response of superconductors to a transverse EM field in the quasiclassical formalism is developed in Ref.~\cite{rai94}, and was applied to superconductors in strong-coupling limit for electron-phonon mediated superconductivity with non-magnetic disorder.
The results reported here for the current response of superconductors, with non-magnetic and paramagnetic impurities, is based on the Keldysh propagator calculated in the linear response limit for an isotropic, spin-singlet (``s-wave'') superconductor using Eliashberg's method~\cite{eli72} of two-frequency analytic continuation~\cite{sau17},
%
\vspace*{-3mm}
\onecolumngrid
\ber
\vJ^R(\vq,\omega)
&=&
2N_f\left(\frac{e^2}{c}\right)
\int dS_{\vp}\vv_{\vp}\left(\vv_{\vp}\cdot\vA(\vq,\omega)\right)
\int\frac{d\varepsilon}{4\pi i}
\times
\nonumber\\
&
\Bigg\{
&
\frac{\pi^2\dpR(\varepsilon,\omega)}
     {\pi^2\dpR(\varepsilon,\omega)^2+(\vq\cdot\vv_{\vp})^2}
\times
\left(1+\frac{\bepsR(\varepsilon+\omega)\bepsR(\varepsilon)+\Delta^2}
             {\sqrt{\Delta^2-\bepsR(\varepsilon+\omega)^2}
              \sqrt{\Delta^2-\bepsR(\varepsilon)^2}}
\right)
\tanh\left(\frac{\varepsilon}{2T}\right)
\nonumber\\
&-&
\frac{\pi^2\dpA(\varepsilon,\omega)}
     {\pi^2\dpA(\varepsilon,\omega)^2+(\vq\cdot\vv_{\vp})^2}
\times
\left(1+\frac{\bepsA(\varepsilon+\omega)\bepsA(\varepsilon)+\Delta^2}
             {\sqrt{\Delta^2-\bepsA(\varepsilon+\omega)^2}
              \sqrt{\Delta^2-\bepsA(\varepsilon)^2}}
\right)
\tanh\left(\frac{\varepsilon+\omega}{2T}\right)
\nonumber\\
&+&
\frac{\pi^2\dpa(\varepsilon,\omega)}
     {\pi^2\dpa(\varepsilon,\omega)^2+(\vq\cdot\vv_{\vp})^2}
\times
\left(1+\frac{\bepsR(\varepsilon+\omega)\bepsA(\varepsilon)+\Delta^2}
             {\sqrt{\Delta^2-\bepsR(\varepsilon+\omega)^2}
              \sqrt{\Delta^2-\bepsA(\varepsilon)^2}}
\right)
\left[
\tanh\left(\frac{\varepsilon+\omega}{2T}\right)
-
\tanh\left(\frac{\varepsilon}{2T}\right)
\right]
\Bigg\} 
\label{eq-Current_Response-paramagentic_impurities}
\,,
\eer
where 
\ber
\dpR 
&=&
-\frac{1}{\pi}
\left[
\sqrt{\Delta^2-\bepsR(\varepsilon+\omega)^2}
+
\sqrt{\Delta^2-\bepsR(\varepsilon)^2}
+
\frac{1}{\tauN}
-\frac{1}{2}
\left(
\frac{1}{\tauSR(\varepsilon+\omega)}
+
\frac{1}{\tauSR(\varepsilon)}
\right)
\right]
\,,
\label{eq-denominator-R}
\\
\dpA 
&=&
-\frac{1}{\pi}
\left[
\sqrt{\Delta^2-\bepsA(\varepsilon+\omega)^2}
+
\sqrt{\Delta^2-\bepsA(\varepsilon)^2}
+
\frac{1}{\tauN}
-\frac{1}{2}
\left(
\frac{1}{\tauSA(\varepsilon+\omega)}
+
\frac{1}{\tauSA(\varepsilon)}
\right)
\right]
\,,
\label{eq-denominator-A}
\\
\dpa 
&=&
-\frac{1}{\pi}
\left[
\sqrt{\Delta^2-\bepsR(\varepsilon+\omega)^2}
+
\sqrt{\Delta^2-\bepsA(\varepsilon)^2}
+
\frac{1}{\tauN}
-\frac{1}{2}
\left(
\frac{1}{\tauSR(\varepsilon+\omega)}
+
\frac{1}{\tauSA(\varepsilon)}
\right)
\right]
\,,
\label{eq-denominator-a}
\eer
\twocolumngrid
where $\bepsR$ is the analytic continuation of the renormalized Matsubara energy given in Eq.~\eqref{eq-bar_epsilonR}, and similarly for $\bepsA$; $1/\tauN$ is the quasiparticle scattering rate by non-magnetic impurities defined in Eq.~\eqref{eq-tauN-tauS}, and
%
\be
\frac{1}{\tauSRA(\varepsilon)}
= 
\frac{1}{\tauS}
(1-\buS^2)
\frac{\Delta^2 - \bepsRA(\varepsilon)^2}
     {\Delta^2(1-\buS^2)^2 - (1+\buS^2)^2\bepsRA(\varepsilon)^2}
\label{eq-tauSinv_vs_epsilon}
\ee
are the retarded and advanced scattering rates contributing to the renormalized gap, $\DeltaRA$, obtained from analytic continuation of Eq.~\eqref{eq-tau2}, with $1/\tauS$ the Born limit scattering rate.

In the absence of paramagnetic impurities Eqs.~\eqref{eq-Current_Response-paramagentic_impurities}-\eqref{eq-denominator-a} reduce to Eqs.~(66)-(68) of Ref.~\onlinecite{rai94} for weak-coupling superconductors.
This was the limit originally studied by Mattis and Bardeen~\cite{mat58} and Abrikosov, Gorkov and Khalatnikov,~\cite{abr59c} for BCS superconductors with non-magnetic impurities.

\subsection{Long-wavelength microwave conductivity}
\vspace*{-3mm}

To leading order in $q v_f \simeq \Delta\,\xi/\lambda\ll\Delta$ we can omit terms of order $(\vq\cdot\vv_{\vp})^2$ in the denominators of Eq.~\eqref{eq-Current_Response-paramagentic_impurities}. The Fermi surface integral is then easily carried out leading to a current density that is a function of the local electric field, $\vJ^R(\vr,\omega)=\sigma(\omega)\,\vE(\vr,\omega)$, with $\vE = i\omega/c\,\vA$. The conductivity is then determined by an integral over the spectrum of charge carriers,

\vspace*{-3mm}
\onecolumngrid
\ber
\sigma(\omega)
=
\frac{2}{3}N_f\,\frac{e^2 v_f^2}{i\omega}\,
\int\frac{d\varepsilon}{4\pi i}
&
\Bigg\{
&
\frac{1}{\dpR(\varepsilon,\omega)}
\times
\left(1+\frac{\bepsR(\varepsilon+\omega)\bepsR(\varepsilon)+\Delta^2}
             {\sqrt{\Delta^2-\bepsR(\varepsilon+\omega)^2}
              \sqrt{\Delta^2-\bepsR(\varepsilon)^2}}
\right)
\tanh\left(\frac{\varepsilon}{2T}\right)
\nonumber\\
&-&
\frac{1}{\dpA(\varepsilon,\omega)}
\times
\left(1+\frac{\bepsA(\varepsilon+\omega)\bepsA(\varepsilon)+\Delta^2}
             {\sqrt{\Delta^2-\bepsA(\varepsilon+\omega)^2}
              \sqrt{\Delta^2-\bepsA(\varepsilon)^2}}
\right)
\tanh\left(\frac{\varepsilon+\omega}{2T}\right)
\\
&+&
\frac{1}
     {\dpa(\varepsilon,\omega)}
\times
\left(1+\frac{\bepsR(\varepsilon+\omega)\bepsA(\varepsilon)+\Delta^2}
             {\sqrt{\Delta^2-\bepsR(\varepsilon+\omega)^2}
              \sqrt{\Delta^2-\bepsA(\varepsilon)^2}}
\right)
\left[
\tanh\left(\frac{\varepsilon+\omega}{2T}\right)
-
\tanh\left(\frac{\varepsilon}{2T}\right)
\right]
\Bigg\} 
\label{eq-Conductivity-paramagentic_impurities}
\nonumber
\,.
\eer
\twocolumngrid

\vspace*{-5mm}
\subsection{Frequency dependent penetration depth}
\label{sec-London_penetration_depth}
\vspace*{-3mm}

An EM field incident from the vacuum penetrates a distance of order the London penetration depth, $\lambda$, into the superconductor. 
For type II superconductors in the limit, $\lambda\gg\xi$, the EM field and screening current penetrate well beyond the vacuum-superconducting interface where surface scattering may modify the bulk excitation spectrum in a region of thickness of order the coherence length, $\xi$, near the interface. 
The relevant wave vectors contributing to the current response are $0\le|\vq|\lesssim\pi/\lambda$.
Thus, to good approximation we can calculate the current response to the penetrating EM field as the response of the homogeneous, bulk superconductor in the long wavelength, $q\rightarrow 0$, limit.
As a result the zero-frequency limit of the imaginary part of the conductivity determines the static London penetration depth, including the effects of disorder, 
$1/\lambda^2=(4\pi/c^2)\lim_{\omega\rightarrow 0}\omega\,\Im{\sigma(\omega)}\equiv 4\pi n_s(T)\,e^2/m^{*}c^2$, where $n_s(T)$ is the superfluid fraction that defines the magnitude of the static supercurrent, and $m^*$ is the normal-state quasiparticle effective mass. 
The effects of impurity disorder on the superfluid fraction of conventional superconductors have been studied in detail~\cite{sau22}.
Although $T_c$, $\Delta(T)$ and the DOS are uneffected by non-magnetic impurities for isotropic superconductors, the presence of condensate flow relative to the random potential is pair breaking, leading to suppression of the superfluid fraction at all temperatures.
Spin-exchange scattering by paramagnetic impurities suppresses $T_c$, $\Delta$ and generates a spectrum of sub-gap quasiparticles even for zero condensate flow. Spin-exchange scattering also contributes to the suppression of the superfluid density as discussed below.

For a microwave EM field the penetration of the field as a function of frequency and temperature is given by
\be\label{eq-Field_penetration_depth}
\frac{1}{\lambda(\omega,T)^2}
= 
\frac{4\pi}{c^2}\,\omega\,\Im{\sigma(\omega)}
\equiv
4\pi\frac{n_s(T,\omega)\,e^2}{m^{*}c^2}
\,.
\ee

In Fig.~\ref{fig-ns_vs_T} we show results for $n_s(T,\omega)$ for $\omega=0.005\,T_c$ for low [$\log(\GammaN/\pi T_{c_0})=-8.29$] and moderate [$\log(\GammaN/\pi T_{c_0})=-0.29$] concentrations of non-magnetic impurities, and for four concentrations of paramagnetic impurities in the limit $\buS=1$, [$\log(\GammaS/2\pi T_{c_0})=-4.45,-3.06,-2.45,-2.05$].
The suppression of the superfluid density at $T=0$ from $n_s\lesssim 1$ to $n_s\approx 0.6$ reflects the pair-breaking effect of a significant concentration of non-magnetic impurities. Additional pair-breaking by dilute concentrations of paramagnetic impurities for $n_s$, and suppression of $T_c$, is also evident.

\begin{figure}[t!]
\includegraphics[width=0.45\textwidth]{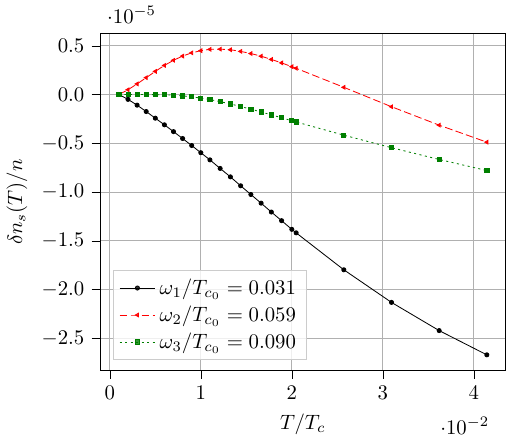}
\caption{Non-monotonic temperature and frequency dependence of the superfluid density for $0<T\lesssim 0.04\,T_c$, $\buS=0.975$, $\GammaS/2\pi T_{c_0}=8.9\times 10^{-6}$ and microwave frequencies covering the position and bandwidth of the sub-gap quasiparticle band near the Fermi level as shown in Fig.~\ref{fig-subgap_DOS-microwave_frequencies}.}
\label{fig-nslope-lowT}
\end{figure}

\begin{figure}[t!]
\includegraphics[width=0.45\textwidth]{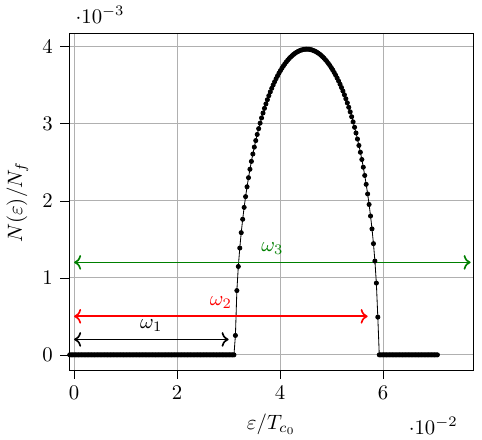}
\caption{The band of sub-gap quasiparticles near the Fermi level generated by a very dilute concentration, $\GammaS/2\pi T_c=8.9\times 10^{-6}$, of paramagnetic impurities with exchange interaction $\buS=0.975$.
Also shown is the reach of microwave frequencies corresponding to the low-temperature superfluid response plotted in Fig.~\ref{fig-nslope-lowT}.}
\label{fig-subgap_DOS-microwave_frequencies}
\end{figure}

What is not visible in Fig.~\ref{fig-ns_vs_T} are non-monotonic anomalies in $n_s(T,\omega)$ at very low temperature that are sensitive to the frequency of the microwave field.
Figure~\ref{fig-nslope-lowT} shows a non-monotonic variation of the superfluid fraction over a narrow temperature range of order $0.04\,T_c$, for three frequencies of the microwave field and $\buS=0.975$. Note that $\delta n_s/n$ is also a non-monotonic function of frequency in this temperature range. The magnitude of the non-monotonic variation is of order $5\,\mbox{ppm}$, which implies similar fractional changes in the penetration of the EM field into the surface of the superconductor, and thus to the resonant frequency of a superconducting resonators.
These low-temperature, low-frequency anomalies in field penentration and resonant frequency originate from a low-concentration of sub-gap quasiparticle bands near, but not overlapping, the Fermi level, within reach of the microwave frequency as show in Fig.~\ref{fig-subgap_DOS-microwave_frequencies}. 
The microwave field produces shifts in the occupation of the sub-gap quasiparticle bands above and below the Fermi level that contribute to the conductivity in Eq.~\eqref{eq-Conductivity-paramagentic_impurities}. For frequency $\omega_2$ shown in Fig.~\ref{fig-subgap_DOS-microwave_frequencies} that spans the positive energy sub-gap band the occupancy of the bands above and below the Fermi level are shifted to effectively suppress the thermal population of states above the Fermi level that would otherwise contribute to thermal backflow, and to increase the occupancy of the negative energy sub-gap band below the Fermi level, both leading to an initial increase in the superfluid fraction for $0<T\lesssim 0.012\,T_c$ as shown in Figure~\ref{fig-nslope-lowT}. 
The sensitivity of the superfluid response is also evident for the frequencies above and below $\omega_2$ in Fig.~\ref{fig-subgap_DOS-microwave_frequencies}.

High-Q superconducting resonators exhibit changes in their resonant frequency in the 1-10 ppm range~\cite{baf25}. Notable are Ta-based CPW resonators operating at $4.8\,\mbox{GHz}$ exhibits a non-monotonic resonant frequency shift of order $\delta f/f \approx 0.5\,\mbox{ppm}$ at low temperatures as shown in Fig. 2b of Ref.~\onlinecite{cro23}. The authors attribute the anomaly to TLS defects. However, the non-monotonic anomaly we find at GHz frequencies results from the sub-gap quasiparticle response to the microwave field. Such states are excluded in the analysis of Ref.~\onlinecite{cro23} because the authors assume Ta is a fully gapped superconductor.

\vspace*{-5mm}
\subsection{Dissipative conductivity}\label{sec-conductivity}
\vspace*{-5mm}

In contrast to the non-dissipative response of the superfluid density, $n_s(T,\omega)$, the dissipative response, $\Re\sigma(\omega,T)$, is sensitive to temperature, frequency, non-magnetic and paramagnetic impurity disorder, for all temperatures and frequencies, including very low-temperatures and microwave frequencies.

\begin{figure}[t!]
\includegraphics[width=\columnwidth]{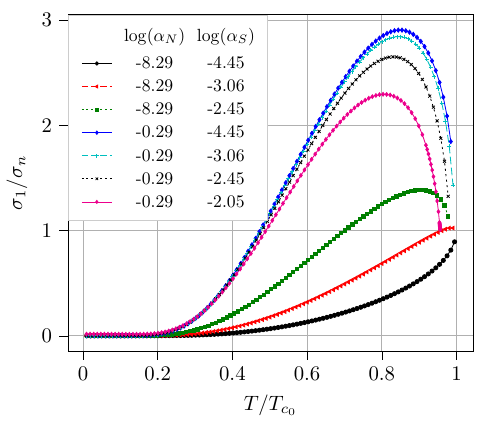}
    \caption{The dissipative microwave conductivity, $\sigma_1(\omega,T)$, as a function of temperature for the same values of magnetic and non-magnetic impurities as those in Fig.~\ref{fig-ns_vs_T}, and $\omega=0.005\,T_c$. The conductivity is normalized by the normal-state conductivity, $\sigma_{n}\equiv\sigma_1(\omega,T_{c})$.}
\label{fig-sigma-T} 
\end{figure}

\begin{figure}[t!]
\includegraphics[width=\columnwidth]{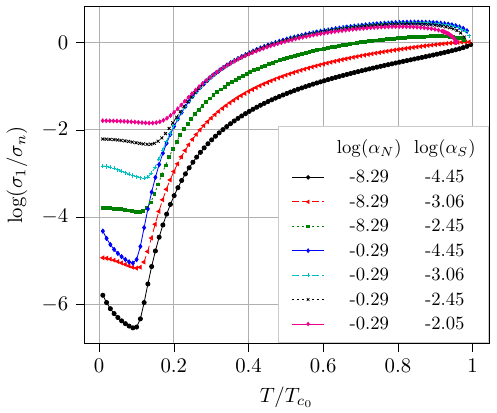}
\caption{The log of the dissipative conductivity, $\sigma_1(\omega,T)$, as a function of temperature for the same parameters as Fig.~\ref{fig-sigma-T}, which highlights the limiting value $\sigma_1(\omega)$ below a temperature of order the bandwidth of the sub-gap impurity band.}
\label{fig-log-sigma-T} 
\end{figure}

Fig.~\ref{fig-sigma-T} shows the dissipative conductivity as a function of temperature for the same parameters for non-magnetic and paramagnetic impurities as those in Fig.~\ref{fig-ns_vs_T}. A key observation is that a large coherence peak just below $T_c$ results from a substantial concentration of non-magnetic impurity disorder, e.g. $\log(\GammaN/2\pi T_{c_0})=-0.29$, a value typical of high-Q niobium SRF cavities~\cite{zar22,zar23,uek25}.
Another key observation is that for a fixed concentration of non-magnetic impurites an additional low concentration of paramagnetic impurities \emph{suppresses} the coherence peak. Thus, the sensitivity of the coherence peak could serve as a signature and measure of very dilute concentrations of paramagnetic impurities. What is not visible in Fig.~\ref{fig-sigma-T} is the low temperature limit of the conductivity. We reveal this regime in Fig.~\ref{fig-log-sigma-T}; the logarithm highlights the saturation of $\Re\sigma(\omega,T)$ below $T\approx 0.2\,T_c$, for the same impurity parameters as those of Fig.~\ref{fig-sigma-T}. The limiting value of $\Re\sigma$ for $T=0$, which is determined by the dissipative response of the small, but finite, density of sub-gap quasiparticles at the Fermi level provides a physical mechanism for the residual resistance inferred from experimental measurements of the limiting $Q$ of superconducting resonators~\cite{rom20}. We highlight these features of the microwave conductivity in the quality factor, $Q$, and resonance frequency shift, $\delta f$, of high-Q superconducting resonators. 

\vspace*{-5mm}
\subsection{Frequency shift, $\delta f$, and $Q$ of cavity resonators}
\label{sec-Q_T1_df}
\vspace*{-3mm}

For transverse fields the current response expressed in terms of the conductivity, $\sigma(\omega)$, and electric field, $\vE(\omega)=i\nicefrac{\omega}{c}\,\vA(\omega)$, is often expressed in terms of the response to the vector potential,
\begin{equation}
\label{eq-J_vs_A}
\vJ^{\text{R}}(\vq,\omega)=-\frac{c}{4\pi}\,K(\omega)\,\vA(\vq,\omega) 
\,.
\end{equation}
The response function, $K(\omega,T)$, is related to the real and imagninary parts of the complex conductivity,
\be\label{eq-K_omega}
K(\omega,T)
=
\frac{1}{\lambda(\omega,T)^2}
-
i\frac{4\pi\omega}{c^2}\,\sigma_1(\omega,T)
\,,
\ee 
where $\lambda(\omega,T)$ is the penetration depth of the EM field into the superconductor (Eq.\eqref{eq-Field_penetration_depth}), and $\sigma_1=\Re\sigma(\omega,T)$ is the contribution to the conductivity that is in-phase with the electric field and thus determines the dissipation of the EM field by Joule losses from un-paired quasiparticles within the region of penetration of the EM field.
For a fully gapped superconductor the Joule dissipation at microwave frequencies is exponentially suppressed at low temperatures leading to a diverging quality factor, $Q(\omega,T)$, for $T\rightarrow 0$.
However, even a very small concentration of paramagnetic impurities with intermediate quasiparticle-impurity exchange interaction leads to dissipation by the microwave field at low temperatures from a band of sub-gap quasiparticles near the Fermi level, and thus a limiting value for $Q$, often interpreted in terms of a phenomenological ``residual'' surface resistance~\cite{rom13a}.
Below we show that a band of sub-gap states with a finite density of states at the Fermi level leads saturation of $Q$ and a residual surface resistance for high-Q cavity resonantors with very low concentrations of paramagnetic impurities.

The field, $\vA(\vr,\omega)$, and the current response, $\vJ^{\text{R}}(\vr,\omega)$, in Eq.~\eqref{eq-J_vs_A} are supplemented by boundary conditions for the vector poential at the vacuum-superconductor interface. The combination allow us to calculate the quality factor and resonant frequency shift of SRF cavities with high precision~\cite{zar23,uek25}.

\vspace*{-5mm}
\subsubsection{Quality factor of high-Q cavity resonators}
\label{sec-Q_vs_T}
\vspace*{-3mm}
 
The quality factor for the fundamental TM mode of a cylindrical SRF cavity at microwave frequencies is related to the current response function, $K(\omega,T)$, by
\begin{equation}
\frac{1}{Q}=\frac{2}{R}\,\Im\left\{\frac{1}{\sqrt{K}}\right\}
\,,
\label{eq-Q}
\end{equation}
where $R$ is the radius of the cavity~\cite{zar23}. 
At microwave frequencies and low temperatures the penetration depth is finite while $\sigma_1(\omega,T)$ is decreasing exponentially with temperature down to very low temperatures as shown in Fig.~\ref{fig-sigma-T}. As a result the quality factor at low temperature is inversely proportional to the dissipative conductivity,
\be
Q(\omega,T)\approx\frac{R}{\lambda_{\omega}}
\left(\frac{c^2}{4\pi\omega\lambda_{\omega}^2}\right)
\,\frac{1}{\sigma_1(\omega,T)}
\,,\quad T \ll T_c
\,,
\ee
where $\lambda_{\omega}=\lambda(\omega,0)\approx\lambda_{\text{L}}$, and $\omega$ is the resonant frequency of the SRF cavity. Thus, the low temperature behavior, including any zero-temperature limit, of $Q$ is determined by the dissipative conductivity $\sigma_1(\omega,T)$ for $T\ll T_c$.
In particular, for high-Q resonantors with very low concentrations of paramagnetic impurities we predict a residual resistance $R_s(0)\propto\sigma_1(\omega,T=0)^{-1}$.
The saturation and zero-temperature limit of $Q$ at low temperatures is shown in Fig.~\ref{fig-lQ-lT} for several decades of very low concentrations of paramagnetic impurities labelled by $\log(\alphaS)$. Note also that the 
upper limit for $Q$ is independent of the concentration of non-magnetic impurities for more than eight orders of magnitude. The maximum value of $Q$ is exponentially sensitive to $\log(\alphaS)$ as is evident in Fig.~\ref{fig-lQ-lT}. 
It is also noteworthy that $Q$ \emph{decreases} as $T$ drops below that corresponding to the maximum $Q$. This non-monotonic temperature dependence is similar to what is observed in high-Q SRF cavities \emph{after} the removal of surface oxides~\cite{rom20}. 
In our theory the non-monotonic temperature dependence is determined by the 
position and width of the sub-gap quasiparticle band 
at the Fermi level. 
It is worth noting that the saturation of $Q$ at low temperature is not limited to pair-breaking by paramagnetic impurities. Embedded two-level tunneling impurities are also capable of creating a band of sub-gap states at the Fermi level, and thus a residual resistance and maximum $Q$ for $T\ll T_c$~\cite{he25}.

\vspace*{-5mm}
\subsubsection{Frequency dependence of $R_s$ and $Q$ for $T\rightarrow 0$}
\label{sec-resistance-Q}
\vspace*{-3mm}

There are two distinct mechanisms for dissipation at microwave frequencies by subgap quasi-particles: 
i) intra-band transitions within each electron or hole impurity band and 
ii) inter-band transitions from hole to particle impurity band and vice versa.  
The former mechanism is active at finite temperature for microwave photons with energies less 
than the  bandwith of the electron or hole impurity band.
Inter-band dissipation requires photons with energies 
spanning the high-energy edge of the hole band below the Fermi level to the low-energy edge of the 
particle band above the Fermi level.

Figure~\ref{fig-iQ-w} shows the inverse quality factor, which is proportional to $R_s$, as a function of 
microwave frequency. The first (second) set of peaks correspond to intra-band (inter-band) transitions.

The distinction between these two processes is revealed in the behavior of inverse quality factor 
as function of temperature for several frequencies in the microwave regime determined by
the sub-gap spectrum in Figure~\ref{fig-subgap_DOS-microwave_frequencies}. 
For microwave frequencies $\omega < 2\omega_1 \approx 0.06\,T_{c_0}$
dissipation results only from intra-band transitions 
for $0<\omega<\omega_2-\omega_1\approx 0.03\,T_{c_0}$. 
The signature of interband transitions is that the microwave dissipation 
decreases monotonically with temperature and vanishes for $T\rightarrow 0$ 
corresponding to a filled hole band and an empty electron band.
By contrast, inter-band absorption onsets at $\omega = 2\omega_1$ and \emph{increases} with 
\emph{decreasing temperature}, reaching a maximum at $T=0$. Both trends are shown Fig.~\ref{fig-iQ-w}.

\begin{figure}[t!]
\includegraphics[width=\columnwidth]{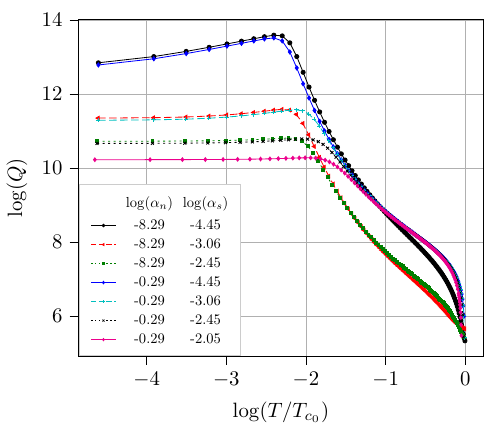}
\caption{Log-log plot of the quality factor as a function of temperature
showing the maximum and zero-temperature limit for $Q$ from a sub-gap 
band of quasiparticles for the same values of magnetic and non-magnetic impurities 
as those in Fig.~\ref{fig-ns_vs_T}.
}
\label{fig-lQ-lT} 
\end{figure}

Microwave absorption measurements on Nb SRF cavities~\cite{rom20} and Ta superconducting circuits~\cite{cro23} 
show evidence consistent with these microwave loss mechanisms in 
their quality factors at low temperatures.
For Nb SRF cavities the quality factor decreases monotonically at low temperatures very similar 
our results shown in Fig.~\ref{fig-lQ-lT}. 
In the case of Ta superconducting circuits the quality factor is also non-monotonic, but 
recovers to a higher value at very low temperatures, similar to the result shown in Fig.~\ref{fig-lQ-T}
which corresponds to a frequency range that includes only intra-band absorption which decreases 
at very low temperatues.

Lastly we note that if the photon energy exceeds the sub-gap bandwidth, but is less than the 
minimum excitation energy required to excite and an occupied state in the hole band to 
and unoccupied state in the particle band
then there is no residual resistivity plateau in $Q(T)$. This case is also highlighted 
in Fig.~\ref{fig-lQ-T}.

\begin{figure}[t!]
\includegraphics[width=\columnwidth]{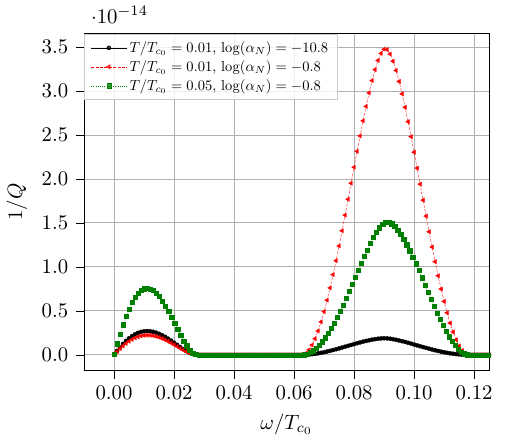}
\caption{The inverse quality factor as a function of frequency for a 
very dilute concentration of paramagnetic impurities with $\GammaS/2\pi T_c=8.9\times 10^{-6}$ 
with exchange interaction $\buS=0.975$. 
The corresponding electron impurity band is shown in Fig.~\ref{fig-subgap_DOS-microwave_frequencies}. 
The first peak in $1/Q$ corresponds to intra-band transitions, while the second peak 
orignates from inter-band transitions between hole-like and particle-like impurity bands. 
}
\label{fig-iQ-w} 
\end{figure}
\vspace*{-5mm}
\subsubsection{Resonant frequency shift of high-Q cavity resonators}
\label{sec-df_vs_T}
\vspace*{-3mm}
 
The resonance frequency of a high-Q cavity generally increases by $1-10$ ppm below $T_c$ due to effective screening of the EM field compared to the normal metal skin depth.
However, there is often a negative frequency shift confined to a narrow temperature region near $T_c$ that was highlighted in detailed measurements reported in Refs.~\cite{baf25}.
A quantitative theory of the small shifts in the resonant frequency of N-doped Nb high-Q resonators, including the negative frequency anomaly, was developed in Refs.~\cite{uek25} and~\cite{zar23}. These authors showed that the quality factor and frequency shift, including the observed temperature dependences, were quantitatively accounted for by the current response of Nb with non-magnetic disorder at levels of disorder below the clean to dirty limit cross-over, i.e. $0.5\lesssim 1/2\pi\tauN T_c \lesssim 0.75$~\cite{uek25}.
They also identified the mechanism responsible for the negative freqency shift anomaly in these resonators; as $T\rightarrow T_c$ from below the London penetration depth diverges and exceeds the normal metal skin depth, leading to a negative frequency shift within a temperature range of width $T_c-T\lesssim 4\times 10^{-3} T_c$ near $T_c$~\cite{zar23}.
Since these studies several groups have reported negative frequency shift anomalies in Nb-based SRF cavities, often with negative shift anomalies extending over a broader temperature range below $T_c$~\cite{ito21,gha23,dha24}.
Here we report new theoretical results for the frequency shift of SRF cavity resonators with non-magnetic impurities and low concentrations of paramagnetic impurities. 

\begin{figure}[t!]
\includegraphics[width=\columnwidth]{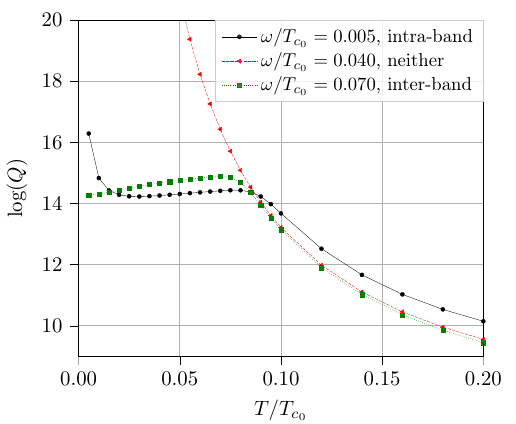}
\caption{
The quality factor as a function of temperature for a very dilute concentration, $\GammaS/2\pi T_c=8.9\times 10^{-6}$, of paramagnetic impurities with exchange interaction $\buS=0.975$ and non-magnetic scattering rate $\alphaN=0.5$. The frequency dependence of $Q$ is determined by the impurity band shown in Fig.~\ref{fig-subgap_DOS-microwave_frequencies}. 
For $\omega=0.07\,T_{c_0}$ (green squares) interband transitions are active and lead to a residual resitivity plateau, and finite value of $Q$ as $T\rightarrow 0$.
However, for $\omega=0.005\,T_{c_0}$ (black circles) intraband transitions are allowed at finite temperature, but 
ultimately suppressed as $T\rightarrow 0$, leading to a plateau followed by increasing $Q$ as $T\rightarrow 0$. 
For $\omega=0.04\,T_{c_0}$ (red triangles) the microwave frequency is insufficient for inter-band transitions, and too large for intraband transitions, in which case the subgap impurity band is \emph{invisible} to microwave photons.
}
\label{fig-lQ-T} 
\end{figure}

The difference in the resonance frequency, $\delta f = f_s-f_0$, of a superconducting resonator, $f_s$, relative to the that of a perfect conductor $f_0$ is determined by the penetration of the EM field into the superconductor, and given by 
\begin{equation}
\frac{\delta f}{f}=-\frac{1}{R}\Re\left\{\frac{1}{\sqrt{K}}\right\}
\,.
\label{eq-df}
\end{equation}

\begin{figure}[t!]
\includegraphics[width=\columnwidth]{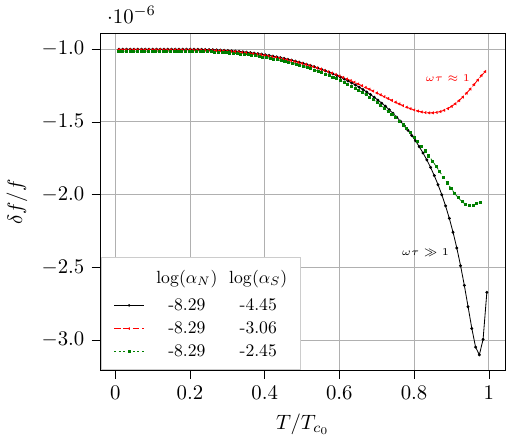}
\caption{The resonant frequency of a superconducting cavity resonator, relative to that for a perfect metal, as a function of temperature, for the nearly clean limit of non-magnetic disorder and a range of concentrations of non-magnetic impurities The temperature width and magnitude of the negative frequency shift anomaly near $T_c$ depends strongly on the ratio $\omega\tau$.}
\label{fig-df-T_clean} 
\end{figure}

Figure~\ref{fig-df-T_clean} shows the impact of paramagnetic impurity scattering on the frequency shift of superconducting resonators for temperatures in the vicinity of $T_c$ for the clean limit for non-magnetic impurities and increasing concentrations of paramagnetic impurities.

\begin{figure}[t!]
\includegraphics[width=\columnwidth]{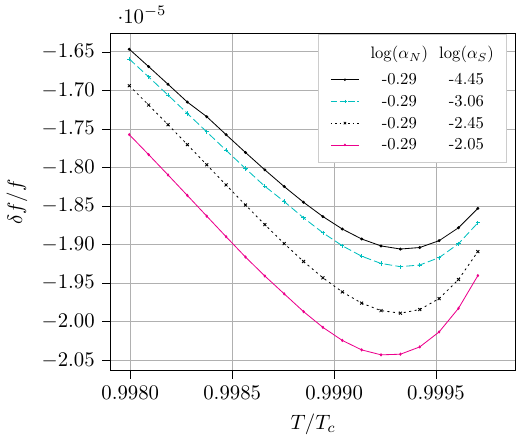}
\caption{The resonant frequency shift of an SRF resonator very near $T_c$ for moderate non-magnetic disorder showing the effect for low concentrations of paramagentic impurities.}
\label{fig-df-T_dirty} 
\end{figure}

An important parameter characterizing the frequency shift anomaly at high temperatures is the ratio of the 
normal-state electron-impurity scattering time, $\tau$, to the period of the microwave field, i.e. $\omega\tau$, where $1/\tau = 1/\tauN + 1/\tauSn$. For $\omega\tau\approx 1$ the frequency shift at $T_c$, $\delta f\vert_{T_c}$, is the same order as the zero-temperature shift, $\delta f\vert_{T=0}$. The corresponding temperature width of the negative frequency shift region is of order, $\Delta T \approx 0.2\,T_c$, which is much wider than anomalous temperature region, $\Delta T \approx 4\times 10^{-4}\,T_c$, for much higher concentrations of non-magnetic impurities characteristic of N-doped Nb SRF resonators~\cite{zar23,uek25}.
The negative frequency shift anomalies shown in Fig.~\ref{fig-df-T_dirty} for $\omega\tau\ll 1$ 
correspond to a concentration of non-magnetic impurities typical of N-doped Nb SRF resonators studied in 
Ref.~\cite{baf25}.

Note (i) the narrow range of the negative frequency anomaly, $\Delta T\approx 6\times 10^{-4} T_c$, and the maximum dip of $\delta f/f_0\approx -0.5\times 10^{-6}$, for the lowest concentration of paramagnetic impurities, $\log\alpha_S = -4.45$, and (ii) the substantial increase in both the width of the anomaly and the magnitude of the dip with increasing concentration of paramagnetic impurities. 
The key point is that the width and magnitude of the negative frequency shift anomaly for $T \lesssim T_c$ are 
sensitive diagnostics of both the scattering rates from non-magnetic impurities, as well as that from 
low concetrations of paramagnetic impurities, in high-Q superconducting resonators.

\vspace*{-5mm}
\section{Conclusion}\label{sec-Conclusion}
\vspace*{-3mm}

Optimizing the performance of superconducting resonators and devices requires identifying and eliminating sources of dissipation and decoherence, often embedded or interfacial defects, or two-level system defets.
This task becomes more challenging as the performance level of the 
device increases. 

We have shown that Joule dissipation from a sub-gap spectrum of un-paired quasiparticles, generated by dilute concentrations of paramagnetic impurities, limits the quality factor of superconduting resonators operating at microwave frequencies at low temperatures.
In particular, a sub-gap band of quasiparticles near the Fermi level provides a microscopic mechanism for zero-temperature residual resistance of high-Q resonators.
Signatures of non-magnetic and paramagnetic impurity disorder are observable in the temperature-dependent shifts of the resonant frequency and the quality factor of high-Q superconducting resonators.

\vspace*{-5mm}
\section{Acknowledgements}
\vspace*{-3mm}
We thank Maria Iavarone, Ruslan Prozorov and John Zasadinskii for discussions and sharing their insights and evidence of sub-gap quasiparticles from STM, PCTS and London penetration depth measurements on superconducting niobium.
This work was supported by the U.S. Department of Energy, Office of Science, National Quantum Information Science Research Centers, Superconducting Quantum Materials and Systems Center (SQMS), under Contract No. 89243024CSC000002,
the Hearne Institute of Theoretical Physics, and the Center for Computational Technology at Louisiana State University.

\appendix

\vspace*{-5mm}
\section{Two-channel Paramagnetic impurity scattering}\label{app-two-channel_T-matrix}
\vspace*{-3mm}

In general paramagentic impurities scatter conduction electrons in both spin-flip (exchange scattering) and 
non-spin flip (potential scattering) channels. Thus, the corresponding matrix element in the s-wave scattering 
limit is
\begin{equation}
\whU=\uN\whone + \uS\,\vec{e}\cdot\widehat{\vec\Sigma}
\,,
\label{eq-UV}
\end{equation}
with $\uS=J\nicefrac{\hbar}{2}\,S$ ($\uN$) the exchange (potential) interaction strength.
Carrying out the configurational averging over the ensemble of randomly oriented impurity spins
the T-matrix based on Eq.~\eqref{eq-UV}, and the analysis to solve the BS equation as described in 
Sec.~\ref{sec-BS_Equation}, is
\ber
\TS 
&=& 
\frac{\uS^2\,\whg\,' + (\uN^2-\uS^2)^2\,\whg + \uN^2\,\whg}
     {(1+\buN^2-\buS^2)^2 + 4\,\buS^2\,\left(\displaystyle{\frac{\tilde\varepsilon_n^2}
                                       {\tilde\varepsilon_n^2+\tilde\Delta^2}}\right)}
\,.
\label{eq-T_S-matrix_two-channel}
\eer
The dependence of the T-matrix on the two propagators, $\whg$ and $\whg\,'$, and the 
renormalized Matsubara frequency and gap amplitude is the same as that in Eq.~\eqref{eq-T_S-matrix}, 
but with both spin-flip and non-spin-flip matrix elements contributing.

The normal-state limit ($\Delta=0$) reduces to s-wave scattering described by two phase shifts,
\be
\TS = \frac{1}{2\pi N_f} \left(\sin^2\delta^+ + \sin^2\delta^-\right)
\,,
\label{eq-T_S-matrix_two-channel_normal-state}
\ee
with $\delta^+$ being the phase shift for conduction electron scattering with electron spin parallel 
to the impurity spin, and similarly, $\delta^-$ is the phase shift for scattering with with the electron 
spin anti-parallel to the impurity spin.
The phase shifts are given by
\be
\delta^{\pm}\equiv \tan^{-1}\left(\buN \pm \buS\right)
\,,
\ee
and correspond to the scattering matrix elements introduced in Refs.~\cite{abr61,rus69a}.

\medskip

Eq.~\eqref{eq-T_S-matrix_two-channel} reduces to Eq.~\eqref{eq-T_S-matrix} 
in the limit of pure exchange scattering, $\uN\rightarrow 0$, and
agrees with Shiba's Eqs.~(3.6) and (3.7) in this limit.
However, our result in Eq.~\eqref{eq-T_S-matrix_two-channel}
differs from Shiba's Eqs.~(4.2) and (4.3) 
which include non-spin-flip scattering.
In comparison to our result in Eq.~\eqref{eq-T_S-matrix_two-channel} Shiba's result 
has an error in the cross terms $\propto \uN^2\,\uS^2$ in the denominator
of the T-matrix. This leads to errors in the bound state spectrum when both 
channels are included.

We checked our result for $\TS$
for the single impurity limit in which the renormalization factors drop out, 
$\tilde\varepsilon_n\rightarrow\varepsilon_n$ and $\tilde\Delta\rightarrow\Delta$.
Analytic continution of $\TS$ to real energies, $i\varepsilon_n\rightarrow\varepsilon + i0^+$,
reveals single-impurity bound state poles at energies,
\ber
\hspace*{-1mm}
\epsilon^{\pm} 
\ns=\ns 
\pm\Delta\cos\left(\delta^+\ns-\ns\delta^-\right)
\ns=\ns 
\pm
\frac{\Delta\,\vert 1-\tilde\uS^2\vert}
     {\sqrt{(1+\tilde\uS^2)^2\ns -\ns 4\tilde\uS^2\,\displaystyle{\frac{\buN^2}{1+\buN^2}}}},
\label{eq-Rusinov_poles}
\eer
with $\tilde\uS\equiv\buS/\sqrt{1+\buN^2}$. 
The result in Eq.~\eqref{eq-Rusinov_poles} also agrees with Rusinov's single impurity 
bound-state energies for s-wave scattering~\cite{rus69a}.

Thus, as is the case for pure exchange scattering, the two-channel T-matrix hosts a pair of 
subgap bound states above and below the Fermi level, with sub-gap energies lying in the range
$0\le|\varepsilon^{\pm}|\le\Delta$ depending on the strengths of the exchange and potential 
matrix elements.

Lastly we note that in the low concentration limit for the paramagnetic impurities 
the dominant contribution to the non-pair-breaking scattering rate contributing to the 
microwave conductivity comes from \emph{non-magnetic impurities}. Only the pair-breaking rate,
$1/\tauSRA(\varepsilon)$ given in Eq.~\eqref{eq-tauSinv_vs_epsilon}
is important for low-concentrations of paramagentic 
impurities as discussed in Sec.~\ref{sec-EM_Response}. 

%
%
\end{document}